%% file: main.tex
\documentclass[ALICE,manyauthors]{cernphprep}
\usepackage[comma,square,numbers,sort&compress]{natbib}

\usepackage{lineno}
\usepackage{xspace}
\usepackage{hyperref}
\usepackage{xcolor}
\usepackage{amsmath}
\usepackage{amssymb}
\usepackage[T1]{fontenc}
\usepackage{orcidlink}

\begin{document}
\input{commands.tex}
%\linenumbers
%%%%%%%%%%%%%%%  Title page %%%%%%%%%%%%%%%%%%%%%%%%
\begin{titlepage}
% the dates below correspond to CERN approval
% please don't touch: EB chairs will take care
\PHyear{2022}       % required, will be obtained from CERN
\PHnumber{126}      % required, will be obtained from CERN
\PHdate{09 June}  % required, will be obtained from CERN
%%%%%%%%%%%%%%%%%%%%%%%%%%%%%%%%%%%%%%%%%%%%%%%%%%%%

%%% Put your own title + short title here:
\title{$\mathbf{\fzero}$ production in inelastic pp collisions at $\mathbf{\sqrt{\textit{s}}}$ = 5.02~TeV}
\ShortTitle{\fzero production in pp collisions}   % appears on left page headers

%%% Do not change the next lines
\Collaboration{ALICE Collaboration\thanks{See Appendix~\ref{app:collab} for the list of collaboration members}}
\ShortAuthor{ALICE Collaboration} % appears on left page headers, do not change

\begin{abstract}

The measurement of the production of \fzero in inelastic pp collisions at \five is presented. This is the first reported measurement of inclusive \fzero yield at LHC energies. The production is measured at midrapidity, \yrange{0.5}, in a wide transverse momentum range, \linebreak \mbox{$0<\pt<16$~\GeVc}, by reconstructing the resonance in the \mbox{\fzero~$\rightarrow\pip\pim$} hadronic decay channel using the ALICE detector. The \pt-differential yields are compared to those of pions, protons and $\phi$ mesons as well as to predictions from the HERWIG~7.2 QCD-inspired Monte Carlo event generator and calculations from a coalescence model that uses the AMPT model as an input. The ratio of the \pt-integrated yield of \fzero relative to pions is compared to measurements in \ee and \pp collisions at lower energies and predictions from statistical hadronisation models and HERWIG~7.2. A mild collision energy dependence of the \fzero to pion production is observed in \pp collisions from SPS to LHC energies. All considered models underpredict the \pt-integrated 2\fzero/(\pip+\pim) ratio. The prediction from the canonical statistical hadronisation model assuming a zero total strangeness content of \fzero is consistent with the data within 1.9$\sigma$ and is the closest to the data. 
The results provide an essential reference for future measurements of the particle yield and nuclear modification in \pPb and \PbPb collisions, which have been proposed to be instrumental to probe the elusive nature and quark composition of the \fzero scalar meson. 

\end{abstract}
\end{titlepage}

\setcounter{page}{2} %please do not remove this line
\input{introduction}

\input{experimentalSetup}
\input{analysis}
\input{resultDiscussion}

\input{conclusions}

%\newpage
%%%%%%%%%%%%%%%%%%%%%%%%%%%%%%%%%
%% end main text 
%%%%%%%%%%%%%%%%%%%%%%%%%%%%%%%%%
%%%%%% acknowledgements - handled by EB chairs 
\newenvironment{acknowledgement}{\relax}{\relax}
\begin{acknowledgement}
\section*{Acknowledgements}
%% add specific acknowledgements here 
%% ...but please don't remove the line below: funding agencies
%% will be acknowledged with a custom tex file handled by EB chairs after Collab Round 2
\input{fa_2022-06-03_Opt_C.tex}
\end{acknowledgement}

%%%%%%%% Bibliography 
\bibliographystyle{utphys}   % Remember we use title in the biblio
\bibliography{main}

%%%%%%%%%%%%%%%%%%%%%%%%%%%%%%%%
% Appendices: yours (if any) + authorlist
%%%%%%%%%%%%%%%%%%%%%%%%%%%%%%%%
\newpage
\appendix

%
%\input{} % put your appendices here (if any)
%

%%%%%% Authorlist - please do not touch: handled by EB chairs 
\section{The ALICE Collaboration}
\label{app:collab}
\input{2022-06-03-Alice_Authorlist_2022-06-03_0_Opt_C.tex}
\end{document}

%% file: commands.tex
%%%%%%%%%%%%%%%%%%%%%%%%%%%%%%%%%%%%%%%%%%%%%%%%%%
% These are some new commands that may be useful 
% for paper writing in general. If other newcommands
% are needed for your specific paper, please feel 
% free to add here. 
%
% The currently available commands are organized in: 
% 1) Systems
% 2) Quantities
% 3) Energies and units
% 4) Detectors
% 5) particle species 
%%%%%%%%%%%%%%%%%%%%%%%%%%%%%%%%%%%%%%%%%%%%%%%%%%

% 1) SYSTEMS 
\newcommand{\pp}           {pp\xspace}
\newcommand{\ppbar}        {\mbox{$\mathrm {p\overline{p}}$}\xspace}
\newcommand{\XeXe}         {\mbox{Xe--Xe}\xspace}
\newcommand{\PbPb}         {\mbox{Pb--Pb}\xspace}
\newcommand{\pA}           {\mbox{pA}\xspace}
\newcommand{\pPb}          {\mbox{p--Pb}\xspace}
\newcommand{\AuAu}         {\mbox{Au--Au}\xspace}
\newcommand{\dAu}          {\mbox{d--Au}\xspace}

% 2) QUANTITIES 
\newcommand{\s}            {\ensuremath{\sqrt{s}}\xspace}
\newcommand{\snn}          {\ensuremath{\sqrt{s_{\mathrm{NN}}}}\xspace}
\newcommand{\pt}           {\ensuremath{p_{\rm T}}\xspace}
\newcommand{\meanpt}       {$\langle p_{\mathrm{T}}\rangle$\xspace}
\newcommand{\ycms}         {\ensuremath{y_{\rm CMS}}\xspace}
\newcommand{\ylab}         {\ensuremath{y_{\rm lab}}\xspace}
\newcommand{\etarange}[1]  {\mbox{$\left | \eta \right | < #1$}}
\newcommand{\yrange}[1]    {\mbox{$\left | y \right | < #1$}}
\newcommand{\dndy}         {\ensuremath{\mathrm{d}N_\mathrm{ch}/\mathrm{d}y}\xspace}
\newcommand{\dndeta}       {\ensuremath{\mathrm{d}N_\mathrm{ch}/\mathrm{d}\eta}\xspace}
\newcommand{\avdndeta}     {\ensuremath{\langle\dndeta\rangle}\xspace}
\newcommand{\dNdy}         {\ensuremath{\mathrm{d}N_\mathrm{ch}/\mathrm{d}y}\xspace}
\newcommand{\Npart}        {\ensuremath{N_\mathrm{part}}\xspace}
\newcommand{\Ncoll}        {\ensuremath{N_\mathrm{coll}}\xspace}
\newcommand{\dEdx}         {\ensuremath{\textrm{d}E/\textrm{d}x}\xspace}
\newcommand{\RpPb}         {\ensuremath{R_{\rm pPb}}\xspace}

% 3) ENERGIES, UNITS
\newcommand{\nineH}        {$\sqrt{s}~=~0.9$~Te\kern-.1emV\xspace}
\newcommand{\seven}        {$\sqrt{s}~=~7$~Te\kern-.1emV\xspace}
\newcommand{\twoH}         {$\sqrt{s}~=~0.2$~Te\kern-.1emV\xspace}
\newcommand{\twosevensix}  {$\sqrt{s}~=~2.76$~Te\kern-.1emV\xspace}
\newcommand{\five}         {$\sqrt{s} = 5.02$~Te\kern-.1emV\xspace}
\newcommand{\twosevensixnn}{$\sqrt{s_{\mathrm{NN}}}~=~2.76$~Te\kern-.1emV\xspace}
\newcommand{\fivenn}       {$\sqrt{s_{\mathrm{NN}}}~=~5.02$~Te\kern-.1emV\xspace}
\newcommand{\LT}           {L{\'e}vy-Tsallis\xspace}
\newcommand{\GeVc}         {Ge\kern-.1emV/$c$\xspace}
\newcommand{\MeVc}         {Me\kern-.1emV/$c$\xspace}
\newcommand{\TeV}          {Te\kern-.1emV\xspace}
\newcommand{\GeV}          {Ge\kern-.1emV\xspace}
\newcommand{\MeV}          {Me\kern-.1emV\xspace}
\newcommand{\GeVmass}      {Ge\kern-.2emV/$c^2$\xspace}
\newcommand{\MeVmass}      {Me\kern-.2emV/$c^2$\xspace}
\newcommand{\lumi}         {\ensuremath{\mathcal{L}}\xspace}

% 4) DETECTORS 
\newcommand{\ITS}          {\rm{ITS}\xspace}
\newcommand{\TOF}          {\rm{TOF}\xspace}
\newcommand{\ZDC}          {\rm{ZDC}\xspace}
\newcommand{\ZDCs}         {\rm{ZDCs}\xspace}
\newcommand{\ZNA}          {\rm{ZNA}\xspace}
\newcommand{\ZNC}          {\rm{ZNC}\xspace}
\newcommand{\SPD}          {\rm{SPD}\xspace}
\newcommand{\SDD}          {\rm{SDD}\xspace}
\newcommand{\SSD}          {\rm{SSD}\xspace}
\newcommand{\TPC}          {\rm{TPC}\xspace}
\newcommand{\TRD}          {\rm{TRD}\xspace}
\newcommand{\TZERO}        {\rm{T0}\xspace}
\newcommand{\TZEROA}       {\rm{T0A}\xspace}
\newcommand{\TZEROC}       {\rm{T0C}\xspace}
\newcommand{\VZERO}        {\rm{V0}\xspace}
\newcommand{\VZEROA}       {\rm{V0A}\xspace}
\newcommand{\VZEROC}       {\rm{V0C}\xspace}
\newcommand{\Vdecay} 	   {\ensuremath{V^{0}}\xspace}

% 4) PARTICLE SPECIES 
\newcommand{\ee}           {\ensuremath{\mathrm{e}^{+}\mathrm{e}^{-}}} 
\newcommand{\pip}          {\ensuremath{\pi^{+}}\xspace}
\newcommand{\pim}          {\ensuremath{\pi^{-}}\xspace}
\newcommand{\kap}          {\ensuremath{\rm{K}^{+}}\xspace}
\newcommand{\kam}          {\ensuremath{\rm{K}^{-}}\xspace}
\newcommand{\pbar}         {\ensuremath{\rm\overline{p}}\xspace}
\newcommand{\kzero}        {\ensuremath{{\rm K}^{0}_{\rm{S}}}\xspace}
\newcommand{\lmb}          {\ensuremath{\Lambda}\xspace}
\newcommand{\almb}         {\ensuremath{\overline{\Lambda}}\xspace}
\newcommand{\Om}           {\ensuremath{\Omega^-}\xspace}
\newcommand{\Mo}           {\ensuremath{\overline{\Omega}^+}\xspace}
\newcommand{\X}            {\ensuremath{\Xi^-}\xspace}
\newcommand{\Ix}           {\ensuremath{\overline{\Xi}^+}\xspace}
\newcommand{\Xis}          {\ensuremath{\Xi^{\pm}}\xspace}
\newcommand{\Oms}          {\ensuremath{\Omega^{\pm}}\xspace}
\newcommand{\degree}       {\ensuremath{^{\rm o}}\xspace}
\newcommand{\fzero}        {\ensuremath{{\rm f}_{0}(980)}\xspace}
\newcommand{\fzeroshort}        {\ensuremath{{\rm f}_{0}}\xspace}
\newcommand{\ftwo}        {\ensuremath{\mathrm{f}_{2}(1270)}\xspace}
\newcommand{\rhozero}        {\ensuremath{\mathrm{\rho}(770)}\xspace}
\newcommand{\fz}        {\ensuremath{\mathrm{f}_{0}}\xspace}

\newcommand{\qqbar}        {\mbox{$\mathrm {q\overline{q}}$}\xspace}
\newcommand{\qqq}        {\mbox{$\mathrm {qqq}$}\xspace}
\newcommand{\qbarqbar}        {\mbox{$\mathrm {\overline{q}\overline{q}}$}\xspace}
\newcommand{\remove}[1] {\sout{#1}}
\newcommand{\kkmolecule}        {\mbox{$\mathrm {K\overline{K}}$}\xspace}
\newcommand{\kstar} {\ensuremath{\mathrm{K}^{*0}}}
\newcommand{\antikstar} {\ensuremath{\overline{\mathrm{K}^{*0}}}} 

\newcommand{\uubar}        {\mbox{$\mathrm {u\overline{u}}$}\xspace}

\newcommand{\ddbar}        {\mbox{$\mathrm {d\overline{d}}$}\xspace}

\newcommand{\ssbar}        {\mbox{$\mathrm {s\overline{s}}$}\xspace}

%% file: introduction.tex
\section{Introduction} \label{sec:introduction}

The conventional picture for the classification of hadrons is based on the constituent quark model introduced in the 1960s~\cite{Gell-Mann:1964ewy}, in which the observed mesons and baryons are described as colourless \qqbar and \qqq bound states, respectively. 
Most of the known observed states fit into the quark model picture. 
At the same time, there are states whose quantum numbers are known but their mass and width have not been measured, and observed resonances whose properties suggest an exotic structure~\cite{Zyla:2020zbs}. 
One remarkable case is that of the light scalar mesons, light-flavoured states with spin zero, positive parity and charge ($J^{PC} = 0^{++}$) and masses below 2 \GeVmass, whose identification represents a long-standing puzzle in particle physics~\cite{Jaffe:1976ig, Jaffe:1976ih, Close:2002zu, Amsler:2004ps, Maiani:2004uc, Klempt:2007cp}. 
From a theoretical point of view, the structure of these states is highly debated~\cite{Zyla:2020zbs}: light scalar mesons could be conventional \qqbar mesons, or compact (qq)(\qbarqbar) structures (tetraquarks), or meson--meson bound states in the form of hadronic molecules, or a superposition of all these components, or glueballs. 

From an experimental point of view, light scalar resonances are typically reconstructed via their dominant decay channels into pseudoscalar mesons (e.g., $\pi\pi$, $\eta\pi$, $\eta\eta$...). The states decaying into pions, in particular, have large characteristic decay widths, of the order of few tens to few hundreds of MeV/$c^{2}$, due to the large available phase space. 
Therefore, the isolation of the particle signals is particularly challenging as broad signals strongly overlap. In addition, for some of the scalar meson states, different decay channels can open up within a short mass interval and distort the line shapes of the nearby resonances. 
 
Among the scalar mesons, the \fzero state is particularly interesting for two reasons. First, despite a long history of experimental and theoretical studies, its nature is still controversial as the properties of the \fzero state are compatible with a conventional \qqbar meson~\cite{Chen:2003za}, a tetraquark~\cite{Achasov:2020aun}, and a \kkmolecule molecular~\cite{Ahmed:2020kmp} structure. 
Secondly, the \fzero represents an interesting probe of the high-density hadronic final state of heavy-ion collisions and in-medium particle formation mechanisms~\cite{Oliinychenko:2021enj}. 

The \fzero couples predominantly to the $\pi\pi$ and \kkmolecule channels and its signal overlaps strongly with the background represented mainly by the f$_0$(500) and the f$_0$(1370), among the scalar mesons. 
An indication in favour of the tetraquark structure of \fzero~\cite{Achasov:2003cn} comes from measurements of the $\phi$ meson radiative decay branching ratios by SND~\cite{Achasov:2000ku}, CMD2~\cite{CMD-2:1999imm}, and KLOE~\cite{KLOE:2002kzf, KLOE:2006vmv} experiments. This is further supported by a recent analysis~\cite{Achasov:2020aun} of the  $J/\psi$ radiative decay data from BESIII~\cite{Achasov:2020aun, BESIII:2015rug}. 
The \fzero is also prominently produced in D$^{+}_{\rm s}$ decays as reported by the E791 collaboration~\cite{E791:2000lzz}, and observed in weak decays of B and B$_{\rm s}$ mesons measured with LHCb~\cite{LHCb:2014ooi, LHCb:2014vbo}. There, the appearance of the \fzero in competition with the $\phi$ meson in these decays could be explained by a large \ssbar component of this state, combined with the fact that the $\rm{c}\rightarrow \rm{s}$ coupling is Cabibbo favoured. In this scenario, the structure of the \fzero would be $\vert\fzero\rangle = \vert (\uubar+\ddbar)\ssbar\rangle/\sqrt{2}$~\cite{Zyla:2020zbs}.
An analysis of the measured couplings of the B and B$_{\rm s}$ mesons to $J/\psi+$\fzero excluded the tetraquark hypothesis~\cite{LHCb:2014vbo}, a conclusion that is however challenged by a different analysis of the same data~\cite{Daub:2015xja}.
Indications that \fzero could be a \kkmolecule molecule come instead from the study of pion--pion and kaon--kaon scattering via non-perturbative QCD methods, which use effective meson-exchange models of the $\pi\pi$ interaction~\cite{Janssen:1994wn, Weinstein:1990gu} and study the \kkmolecule interaction for coupled and single channels in chiral effective theory~\cite{Xiao:2019lrj,Ahmed:2020kmp}.

In addition to measuring the production rates and branching fractions of \fzero in $\phi$ and heavy-flavour decays, several authors~\cite{Maiani:2004uc, Maiani:2006ia, ExHIC:2017smd, Gu:2019oyz} have proposed to investigate its nature by using heavy-ion collisions and exploiting the unique production (and decay) environment accessible in these reactions.
In high-energy heavy-ion collisions, two extreme states of matter are reached one after the other. If enough energy is deposited in the collision region, the state of deconfined strongly interacting matter called quark--gluon plasma (QGP) is produced and expands as a nearly perfect liquid
until the temperature reaches the pseudo-critical value of $\approx 155$~MeV~\cite{Bazavov:2018mes} and a transition to confined QCD matter takes place. A hot (T $\approx$ 100--150 MeV) and dense gas of interacting hadrons is formed in which resonances decay and particles interact (pseudo)elastically until they decouple. 
At the LHC, the system produced in \PbPb collisions decouples after about 10 fm/$c$~\cite{ALICE:2011dyt} and the production of hadronic resonances with lifetimes of the order of 1 to 10 fm/$c$ is studied to characterise the hadronic stage of the collision~\cite{ALICE:2018qdv, ALICE:2019xyr, ALICE:2018ewo}.
With its width between 10 and 100 MeV/$c^{2}$ and a corresponding lifetime of $\approx$~5--10 fm/$c$,
the \fzero is a probe for the dense hadron gas formed in the late stage of heavy-ion collisions~\cite{Oliinychenko:2021enj}.

Measurements of the nuclear modification factor~\cite{Maiani:2006ia}, the particle yield per event~\cite{ExHIC:2017smd}, and the elliptic flow coefficient~\cite{Gu:2019oyz} have been suggested to provide insights into the internal structure of the \fzero.
Models of hadron formation via recombination (coalescence)~\cite{Fries:2003vb,Minissale:2015zwa,Plumari:2017ntm} of quarks in the quark--gluon plasma that have been successful in describing LHC data, indicate that the \fzero production in the intermediate transverse momentum range (2 $<\pt<$ 5 \GeVc) is sensitive to the number of constituent quarks. 
Theory calculations based on a coalescence model~\cite{ExHIC:2017smd} show that the \pt-integrated production of \fzero in central heavy-ion collisions at LHC energies is expected to be two orders of magnitude lower if the state has a tetraquark structure compared to the results for a non-exotic diquark structure \qqbar, or a hadronic molecule configuration. 
On the other hand, the production of a tetraquark state would be enhanced in heavy-ion collisions with respect to \pp collisions at the same energy in the $\approx$ 2--6~\GeVc momentum range~\cite{Maiani:2004uc, Maiani:2006ia}.  Measurements of the nuclear modification factor~\cite{Maiani:2004uc, Maiani:2006ia} or of the \pt-dependent yield ratio of the \fzero to particles with different (but established) quark content could therefore shed light on the nature of the state. 
The authors of~\cite{Gu:2019oyz} also suggest that the azimuthal production asymmetry in the \fzero momentum distributions, quantified by the elliptic flow coefficient, could be sensitive to the number of constituent quarks in the kinematic range in which hadron formation occurs predominantly via quark recombination (coalescence). 
A measurement of the \fzero production in \pp collisions is necessary for the determination of the nuclear modification factor and constitutes a reference for the study of the particle production in heavy-ion collisions. 

In this letter, the first measurement of the inclusive production of \fzero in inelastic pp collisions at the LHC is reported. 
To provide a baseline for studies in heavy-ion interactions, the data using collisions at \five were analysed, corresponding to the centre-of-mass energy per nucleon pair of the \pPb and \PbPb data samples collected during the LHC Run 2. Measurements of \fzero in \pPb and \PbPb collisions at this energy will be the subject of future publications.
The production of \fzero is measured at midrapidity, $\vert y\vert  < 0.5$, in a broad transverse momentum range between 0 and 16 \GeVc. 
An overview of the ALICE experimental setup is given in Sec.~\ref{sec:experiment}, followed by a description of the analysis strategy in Sec.~\ref{sec:analysis}. This includes details on the data sample, the \fzero signal reconstruction, the yield extraction and corrections, and the systematic uncertainty estimation. Results are discussed in comparison to lower energy data and theoretical models in Sec.~\ref{sec:results}, while in Sec.~\ref{sec:Conclusions} the conclusions are summarised.

%% file: experimentalSetup.tex
\section{Experimental setup} \label{sec:experiment}
The experimental setup and details on the performance of the ALICE detector are described in Refs.~\cite{ALICE:2014sbx, Aamodt:2008zz}. The ALICE detector consists of a central barrel with a set of detectors devoted to the reconstruction and identification of the charged particles, a forward muon spectrometer and a set of backward and forward systems for triggering and event characterisation purposes. The central barrel detectors are located inside a solenoidal magnet that provides a magnetic field of 0.5 T.
The main detectors employed for the analysis presented in this work are the \VZERO, the Inner Tracking System (ITS), the Time Projection Chamber (TPC), and the Time-of-Flight detector (TOF). 
The \VZERO consists of two scintillator arrays placed on both sides of the interaction point covering the pseudorapidity regions $2.8< \eta < 5.1$ (\VZEROA) and $-3.4 < \eta < -1.7$ (\VZEROC), respectively. The \VZERO provides the minimum bias trigger of the experiment and is used for suppressing beam-induced background at the offline analysis level. 
The position of the collision vertex and the tracks of charged particles are reconstructed in the central barrel using the ITS and the TPC. The ITS is a high-resolution tracker that consists of six cylindrical layers of silicon detectors. 
The TPC is a large cylindrical drift detector covering a radial distance of 85 $< r <$ 247 cm from the beam axis and having longitudinal dimensions of about -250 $<z<$ 250 cm. 
The TOF is a large area array of multigap resistive plate chambers, placed at a radius of about 370--399 cm from the beam line. 
In the central barrel, charged particles can be identified via measurements of their specific energy loss, \dEdx, provided by the TPC with a resolution of 5$\%$, and via their time-of-flight measured by the TOF with a resolution of about \mbox{80 ps}. 

%% file: analysis.tex
\section{Data analysis}
\label{sec:analysis}
The measurement of \fzero production is performed using a sample of minimum bias \pp collision events at a centre-of-mass energy of \five, collected in the years 2015 and 2017. 
The minimum-bias trigger requires at least one hit in both \VZEROA and \VZEROC detectors~\cite{ALICE:2013axi}. The integrated luminosity after trigger selection is $\approx21.8$~nb$^{-1}$.
Events are selected for the analysis if the position of the reconstructed collision vertex along the beam axis is located within 10 cm from the nominal interaction point. 
To reduce the pileup caused by multiple interactions in the same bunch crossing, a criterion based on the offline reconstruction of multiple primary vertices in the two innermost layers of the ITS, namely the Silicon Pixel Detector (SPD) is applied~\cite{ALICE:2014sbx}. The rejected events account for less than 1\% of the total events.
After applying these selection criteria, $\approx9.14\times10^{8}$ collision events have been analysed.

The \fzero resonance signal is reconstructed via its decay into a pair of oppositely charged pions, $\fzero\rightarrow\pip\pim$. 
This requires the reconstruction, selection and identification of pion tracks in the central barrel of ALICE. To ensure a uniform detector acceptance, only charged tracks with \pt $>$ 0.15 \GeVc and pseudorapidity $|\eta|~<$ 0.8 are considered for the analysis. 
Track selection criteria are applied to the charged tracks as in previous works~\cite{ALICE:2019xyr,ALICE:2018qdv} to ensure a good quality of the reconstruction. To this end, each track in the TPC is required to have crossed at least 70 readout pad rows out of a maximum possible 159. 
To reduce the contamination from secondary particles, tracks are accepted if their distance of closest approach to the collision vertex in the longitudinal ($d_{\rm z}$) and transverse ($d_{\rm {xy}}$) directions satisfy $d_{\rm z}<2$ and $d_{\rm{xy}}<0.0105$ $+0.0350\times p_{\rm T}^{-1.1}$, where \pt and distance are in units of \GeVc and \,cm, respectively.
 
The identification of pions is performed using the TPC and the TOF detectors and criteria based on the difference between the measured and expected signals for a given particle hypothesis, divided by the resolution ($\sigma_{\rm{TPC}}$, $\sigma_{\rm{TOF}}$). 
In the TPC, charged particles are identified as $\pi$ if the measured \dEdx is compatible with the expected pion mean specific energy loss within two standard deviations (2$\sigma_{\rm{TPC}}$) over the entire momentum range. 
If a measurement of the particle time-of-flight by the TOF is available, a TOF-based 3$\sigma_{\rm{TOF}}$ selection criterion is applied on top of the TPC-based one, over the measured momentum range.

\subsection{Raw yield extraction}
\begin{figure}[t]
\begin{center}
\includegraphics[width=0.5\textwidth]{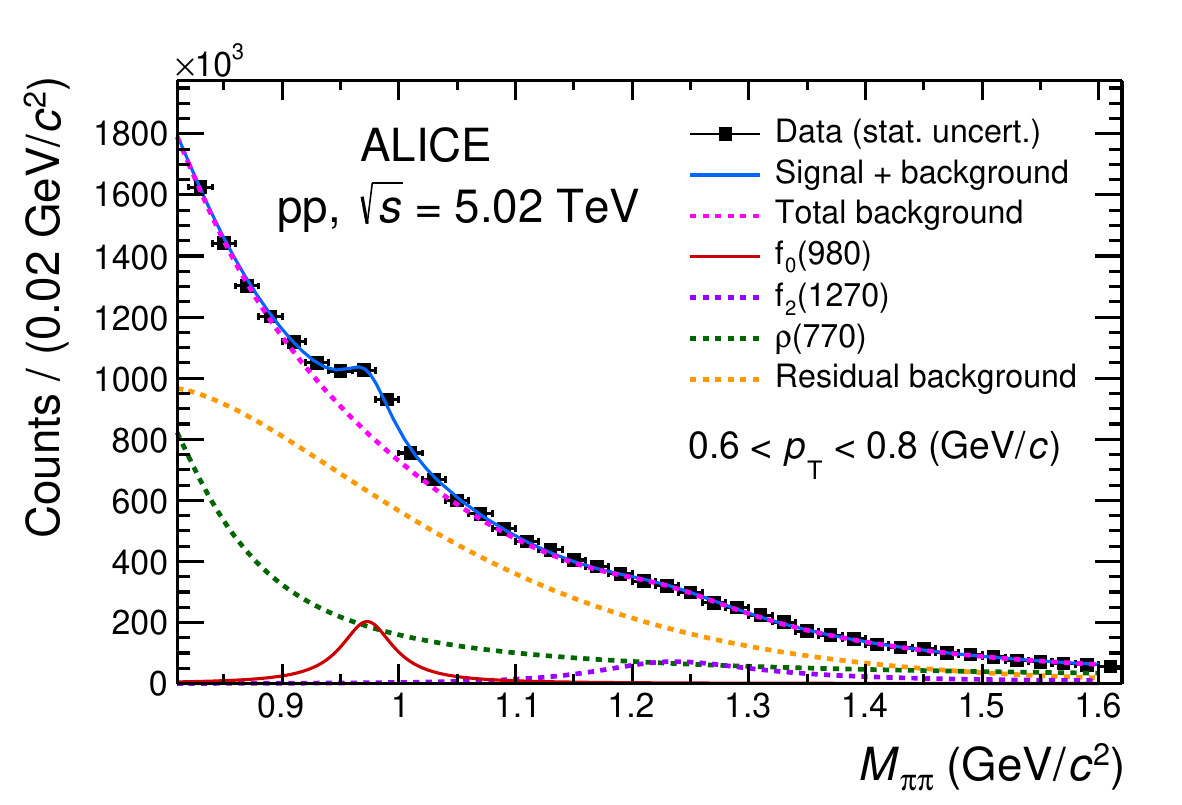}%scale=0.39
\includegraphics[width=0.5\textwidth]{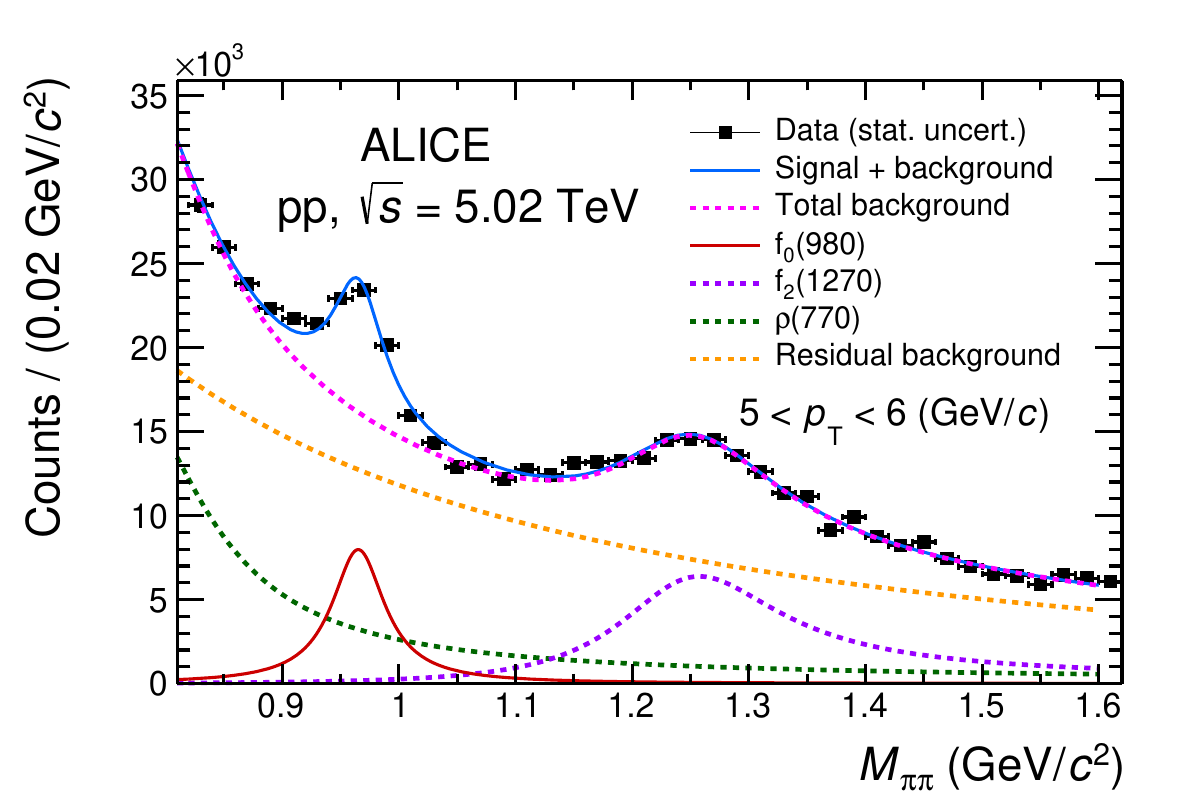}
\caption{Left (right) plot shows the invariant mass distribution of \ensuremath{\pi^{+}}\ensuremath{\pi^{-}} pairs after like-sign background subtraction in low (high) transverse-momentum interval in \pp collisions at \five in $\vert y\vert < 0.5$. Solid blue curves represent fits with the function shown in Eq.~\ref{eq1} and a residual background shown in Eq.~\ref{eq3}.  Solid red curve represents \fzero signal while other dashed curves represent the background contributions from \rhozero, \ftwo and residual background.}
\label{fig:invmass}
\end{center}
\end{figure}

The \fzero resonance signal is reconstructed via an invariant mass analysis by combining oppositely-charged pions within the same event into pairs and imposing the pair to have a rapidity within the range $|y|~<$ 0.5. 
To remove the combinatorial background, the like-sign method is employed. The same-charge pion tracks from the same event are combined into $\pip\pip$ and $\pim\pim$ pairs. The total like-sign invariant mass distribution is calculated as the geometric mean of the positively-charged and negatively-charged pair distributions, as $2\sqrt{N^{++}N^{--}}$, where $N^{++}$ and $N^{--}$ are the number of $\pip\pip$ and $\pim\pim$ pairs, respectively. 
The $\pip\pim$ and like-sign background invariant mass distributions are extracted for various intervals of the pair \pt, and for each of these, the like-sign background is subtracted from the unlike-sign pair distribution.
After the subtraction of the combinatorial background, the \fzero signal peak, sitting on the right-hand tail of the broad \rhozero meson signal, is visible on top of a residual background. Two examples of the \pip\pim invariant mass distributions after combinatorial background subtraction are shown in Fig.~\ref{fig:invmass} for a low-\pt and for a high-\pt interval. 
With increasing \pt, the significance of the \ftwo resonance signal increases and the broad \ftwo peak becomes visible on the right side of the \fzero signal. 
The residual background originates from correlated $\pip\pim$ pairs from mini-jets and from misidentified particles. 
The main contributions to the correlated background arise from the decay of the \rhozero and the \ftwo resonances into oppositely-charged $\pi$ pairs.
In order to extract the \fzero yields in each \pt interval, the distributions are fitted in the invariant mass interval \mbox{0.8 $< M_{\pi\pi}<$ 1.6 \GeVmass} with a function that is the sum of three relativistic Breit-Wigner functions (rBW) describing the \rhozero, \fzero and \ftwo signals~\cite{ALICE:2018qdv,STAR:2002caw,ALICE:2015nbw}, and a residual background. Since the resolution on the invariant mass is negligible with respect to the natural width of the considered resonances, the resonance shape can be modelled with a rBR with no need for any additional Gaussian smearing to account for detector resolution effects.
Each of the rBW functions is defined as

\begin{equation}
{\rm{rBW}}(M_{\pi\pi}) = \frac{AM_{\pi\pi}\Gamma(M_{\pi\pi}) M_{0}}{(M_{\pi\pi}^2-M_0^2)^2+M_0^2\Gamma^2(M_{\pi\pi})}
\label{eq1}
\end{equation}
where $\Gamma(M_{\pi\pi})$ is given by

\begin{equation}
\Gamma(M_{\pi\pi})= \biggl[ \frac{(M_{\pi\pi}^2 - 4m_{\pi}^2)}{(M_{0}^2 - 4m_{\pi}^2)} \biggl] ^{(2J+1)/2} \times  \frac{\Gamma_{0}M_{0}}{M_{\pi\pi}}.
\label{eq2}
\end{equation}

Here, $A$ is the normalisation constant, $M_{0}$ and $\Gamma_{0}$ are the rest mass and width of the resonance, $m_{\pi}$ is the charged pion mass and the spin is $J=0$ for \fzero, $J =1$ for \rhozero and $J=2$ for \ftwo. The shape of the residual background resembles that of a Maxwell-Boltzmann distribution and therefore it is fitted with a similar functional form $f_{\rm bg}(M_{\pi\pi})$
\begin{equation}
f_{\rm bg}(M_{\pi\pi})= B \sqrt{(m_{\pi\pi}-m_{\rm{cutoff}})^n}~C^{3/2}~\exp{[-C(m_{\pi\pi}-m_{\rm{cutoff}})^n]},
\label{eq3}
\end{equation}
where $B$ is the normalisation constant and $m_{\rm{cutoff}}$ is the low-mass cutoff expected to be equal to the rest mass of the $\pip\pim$ pair. This function was proven to provide a good description of the residual background in previous analyses~\cite{ALICE:2018ewo}.
The residual background term takes also into account any possible additional background from f$_0$(500) and f$_0$(1370), which have not been added to the signal model due to the large indetermination\footnote{The f$_0$(500) width ranges from 400 to 700 MeV, the f$_0$(1370) width ranges from 200 to 500 MeV~\cite{Zyla:2020zbs}} on the broad width parameter of these states.

For the extraction of the particle yields, the fits are performed with the following configuration of the fit parameters. The mass and the width of the \rhozero, and the width of the \ftwo are fixed to their vacuum values, $m_{\rho} = 775.26$~\MeVmass, $\Gamma_{\rho} = 149.1$~\MeVmass, and $\Gamma_{f_{2}} = 186.7$~\MeVmass~\cite{Zyla:2020zbs}. The width of the \fzero is fixed to the average value of the range reported in Ref.~\cite{Zyla:2020zbs} that corresponds to $\Gamma_{f_{0}} = 0.055$~\GeVmass. The masses of the \fzero and the \ftwo, as well as the $m_{\rm{cutoff}}$, $C$ and $n$ parameters of $f_{\rm bg}$ are left free.
The fit parameter configuration has been varied to take into account possible imperfections in the description of the background and signal shapes, as discussed in Section~\ref{sec:syst}.
In Fig.~\ref{fig:invmass}, the fit result to the invariant mass distribution of \ensuremath{\pi^{+}}\ensuremath{\pi^{-}} pairs after like-sign background subtraction is shown for two \pt intervals, namely $0.6<\pt<0.8$~\GeVc and $5<\pt<6$~\GeVc.

\subsection{Yield corrections}
In order to obtain the \fzero production yield per unit of rapidity and \pt per inelastic event $\big(\frac{1}{N_{\rm INEL}}\frac{{\rm d}^2N}{{\rm d}y{\rm d}\pt}\big)$, several correction factors are applied to the raw yields obtained from the fit procedure in each \pt interval according to the following formula
\begin{equation}
\frac{1}{N_{\rm INEL}}\frac{{\rm d}^2N}{{\rm d}\pt{\rm d}y} = \frac{1}{N_{\rm evt}}\frac{N_{\fzero\rightarrow\pi\pi}}{{\rm \Delta}\pt {\rm \Delta}y}\frac{\epsilon_{\rm{trig}}~\epsilon_{\rm{vtx}}}{A\times\epsilon_{\rm{rec}}}\frac{f_{\rm{sig}}}{{\rm BR}}.
\end{equation}
Here, $N_{\fzero\rightarrow\pi\pi}$ is the \fzero raw yield measured in a given rapidity (${\rm \Delta}y$) and transverse momentum (${\rm \Delta}$\pt) interval, $N_{\rm evt}$ is the number of collision events that satisfy the selection criteria. The minimum-bias trigger efficiency, the vertex reconstruction efficiency and the signal loss correction factor are represented by $\epsilon_{\rm{trig}}$, $\epsilon_{\rm{vtx}}$ and $f_{\rm{sig}}$, respectively.
The branching ratio correction amounts to BR = (46 $\pm$ 6)\%~\cite{Stone:2013eaa} assuming dominance of $\pi\pi$ and KK channels.
The yields of \fzero are normalised to the number of inelastic pp collisions with a trigger efficiency correction, \mbox{$\epsilon_{\rm{trig}}$ = 0.757 $\pm$ 0.019}~\cite{ALICE:2016mfm, Loizides:2017ack}, which takes into account the efficiency of the \VZERO-based trigger to select inelastic events. The vertex reconstruction efficiency in \pp collisions at \five is found to be $\epsilon_{\rm{vtx}}$ = 0.958~\cite{ALICE:2019xyr}.
The $A\times\epsilon_{\rm{rec}}$ factor corrects for the detector acceptance times the \fzero reconstruction efficiency and is evaluated using a detailed Monte Carlo simulation of the ALICE detector geometry, material, and response. The \pp collision events are simulated using the PYTHIA 8 event generator~\cite{Skands:2014pea} with the addition of the \fzero signals. The generated particles in the simulation are propagated through the detector using GEANT3~\cite{Brun:1994aa}. The $A\times\epsilon_{\rm{rec}}$ is calculated in the rapidity range $\vert y \vert < 0.5$ as a function of \pt and is defined as the ratio of the number of reconstructed and generated \fzero. The reconstruction of \fzero in the simulation is performed using the same event and track selection criteria as employed for the analysis of the data.
\\The signal loss correction factor, $f_{\rm{sig}}$, accounts for the fraction of \fzero signal lost due to trigger inefficiencies and can be determined as a function of \pt using Monte Carlo simulations. Because a simulation with injected \fzero signals may not lead to a realistic estimate of this correction factor, the correction is taken to be the same as for the $\phi$ meson at the same collision energy. The earlier analysis in~\cite{ALICE:2019xyr} showed that this correction does not depend significantly on the particle mass for resonances decaying strongly into two charged particles. 
This factor ranges between 1.07 for \mbox{0 $<$ \pt $<$ 0.2~\GeVc} and 1 for \mbox{\pt $>$ 2.5~\GeVc}.

\subsection{Systematic uncertainties} \label{sec:syst}
\begin{table} [!hpt]
                \centering 
                \caption{Contributions to the relative systematic uncertainty of the \pt-dependent yield of \fzero in \pp collisions at \five. The uncertainties are given for the lowest and the highest \pt intervals of the measured spectrum as well as for one intermediate \pt interval. The total uncertainty is obtained as the sum in the quadrature of the individual contributions. Values are expressed in percentage (\%).
                 \label{tab:sys}}
                \begin{tabular}{l r r r }
                \hline
                \hline
                \textbf{Source of uncertainty}  & \multicolumn{3}{c}{{\pt (\GeVc)}}\\ 
                & 0--0.2 & 4--4.5 & 12--16 \\
                \hline
                Yield extraction        &7.1\% &8.8\% & 15.3\% \\
                Track selection         &9.3\% &2.2\% &2.1\%   \\
                Global tracking efficiency  &2\% &4\% &4\%     \\    
                Particle identification &6.8\% &1.5\% &6\%     \\
                Event selection         &7.6\% &2.1\% &3.3\%   \\
                Material budget         &5.2\% &0\% &0\%       \\
                Hadronic interaction    &3.4\% &0\% &0\%       \\
                \hline
                Total &16.8\% &10.2\% &17.4\%  \\
                \hline
                \hline
                \end{tabular}
\end{table}

The sources of systematic uncertainty in the measurement of the \fzero yields are summarised in Table~\ref{tab:sys}. These include yield extraction, track and event selection, global tracking efficiency, particle identification, the knowledge of the ALICE material budget, and that of the hadron interaction cross section in the detector material. The estimated values of the uncertainties are reported in Table~\ref{tab:sys} for low, intermediate and high-\pt intervals. The systematic uncertainty associated with the yield extraction arises from the fit procedure and is determined by varying the fitting range as well as the signal and the background fit parameters. In particular, the width of the \fzero was varied 
by sampling the range from 10 to 100 \MeVmass~given in~\cite{Zyla:2020zbs} with 15 variations and the width of the \ftwo was varied within $\pm 7.5$~\MeVmass that corresponds to a $\pm3\sigma$ range of the width value reported in~\cite{Zyla:2020zbs}. These variations result in the largest contribution to the uncertainty on the yield extraction.
The uncertainties due to the yield extraction are \pt dependent and vary from 7.1\% in the lowest \pt interval, to 15.3\% in the highest \pt interval of this analysis. 
The systematic uncertainty due to the track selection is evaluated by varying a single track selection criterion at a time in both data and simulation, and by repeating all the steps of the analysis. This contribution ranges from 9.3\% to 2.1\% from low to high \pt.
The difference in the efficiency of the matching of TPC tracks to ITS clusters (global tracking efficiency) between data and simulations results in a contribution to the systematic uncertainty of 2--4\% depending on \pt. 
The systematic uncertainty associated with the particle identification is due to an imperfection in the description of the \dEdx in the TPC-based $n\sigma$ selection in the Monte Carlo simulation as compared to data. The $n\sigma$ selection is varied in data and simulation simultaneously to a 3$\sigma_{\rm TPC}$ particle identification criterion and results in a \pt-dependent relative systematic uncertainty of 1.5--6.8\%. 
The choice of the event selection criteria leads to a systematic uncertainty of 2.1--7.6\%. 
The systematic uncertainty associated with the signal loss correction is estimated by comparing the correction for $\phi$ mesons, used as a proxy for \fzero, with that of other light-flavour hadrons and is found to be lower than 1\%.
Finally, the uncertainty on the knowledge of the ALICE material budget and that of the hadron interaction cross section in the detector material leads to a systematic uncertainty lower than 5.3\% and 3.4\%, respectively~\cite{Acharya:2018qsh,Aamodt:2008zz,Adam:2015bg}. 
The total relative systematic uncertainty is obtained as the sum in the quadrature of these contributions.

%% file: resultDiscussion.tex
\section{Results and Discussion} 
\label{sec:results}

\begin{figure}[htb]
\begin{center}
\includegraphics[scale=0.5]{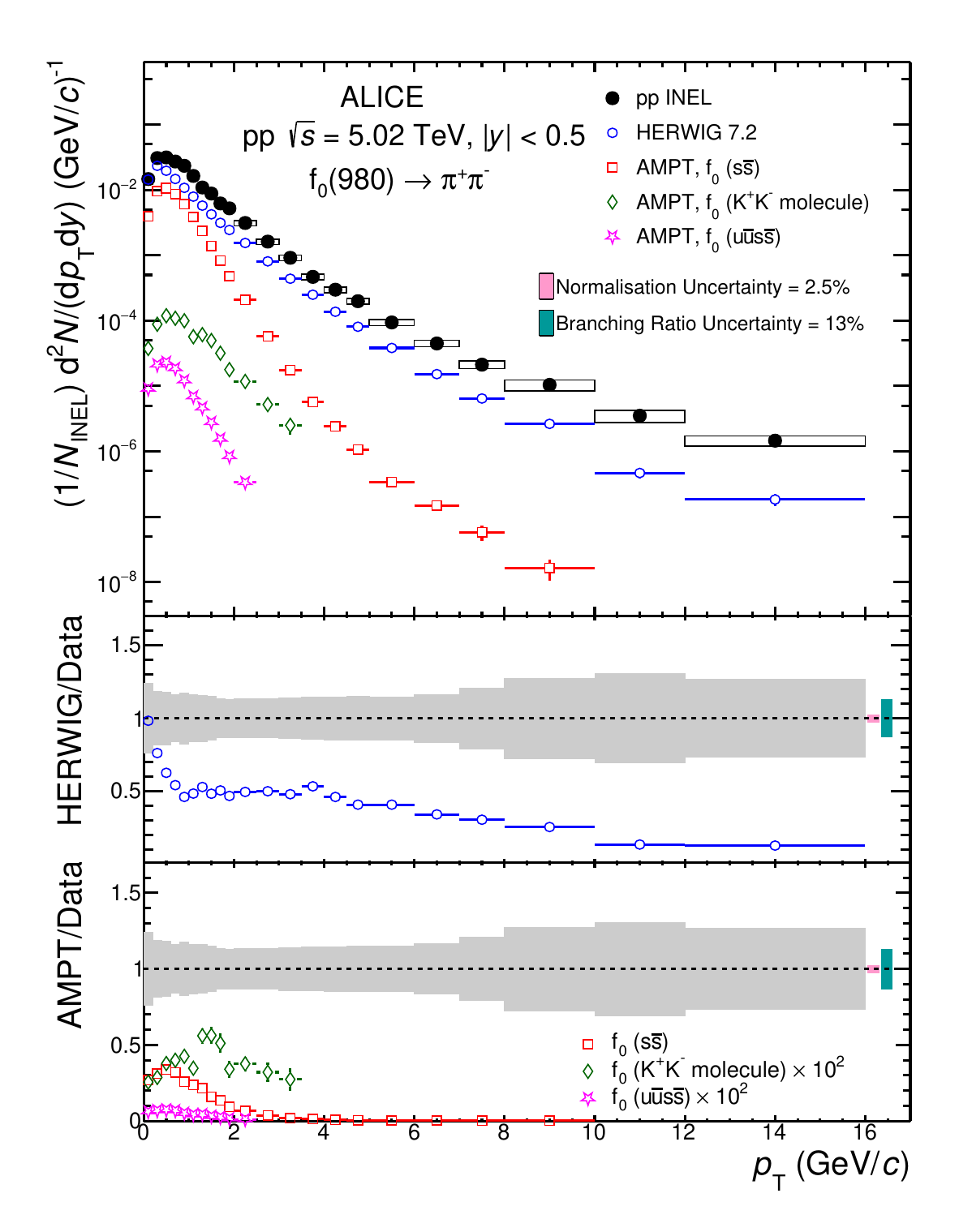}
\caption{The \pt-differential yield of \fzero in \pp collisions at \five is compared with predictions from the HERWIG~7.2 event generator~\cite{Bellm:2015jjp,Bahr:2008pv} and with a coalescence calculation~\cite{Gu:2019oyz} based on the AMPT model~\cite{Lin:2004en}. The statistical and systematic uncertainties on data (full black markers) are shown as bars and boxes, respectively. The middle and bottom panels show the model to data ratios. The grey boxes at unity represent the sum in quadrature of the statistical and systematic uncertainty on the data. The ratios of $\rm{u\bar{u}s\bar{s}}$ tetraquark and $\rm{K^+K^-}$ molecule configurations from AMPT model predictions to data are multiplied by 100 to improve visibility. In all three panels, the uncertainties associated with the models are statistical ones.}\label{fig:ptspectrum}
\end{center}
\end{figure}

The \pt-differential yield of \fzero for $|y|~<~0.5$ in inelastic \pp collisions at \five is shown in the upper panel of Fig.~\ref{fig:ptspectrum}. The measurement spans a wide \pt range from 0 to 16 \GeVc. 

The normalisation and branching ratio relative uncertainties on the yields are independent of \pt and amount to 2.5\% and 13\%, respectively~\cite{Stone:2013eaa,Loizides:2017ack}. 

At present, most of the Monte Carlo generators commonly employed to simulate pp collisions do not implement the generation of \fzero in their default configurations. 
One notable exception is the HERWIG~7.2 event generator~\cite{Bellm:2015jjp,Bahr:2008pv}. HERWIG~7.2 is a QCD-inspired Monte Carlo event generator that includes processes like initial and final state QCD radiation, a description of the underlying event via an eikonal multiple parton--parton interaction model, and a cluster hadronisation model for the formation of hadrons from the quarks and gluons produced in the parton shower. The default hadronisation and shower parameters are tuned to e$^{+}$e$^{-}$ data~\cite{Bahr:2008pv} with the addition of a tune for multi-parton processes based on the minimum bias LHC data~\cite{Bellm:2015jjp}.
To allow for the comparison, model calculations have been performed in the same \pt intervals of the data. 
As shown in the HERWIG/Data ratio reported in the middle panel of Fig.~\ref{fig:ptspectrum}, HERWIG underestimates the measured yields by a factor of about two for $1 < \pt < 4$~\GeVc but reproduces at least qualitatively the shape of the \pt spectrum in this range. 
At \mbox{\pt $\lesssim 0.5$~\GeVc}, the model is consistent with data within uncertainty but the \pt dependence is not described. 
At $p_{\rm T}\geq 4$\,\GeVc, HERWIG is not able to reproduce the data neither qualitatively nor quantitatively.  

The data are also compared to a recent coalescence calculation~\cite{Lin:2004en,Gu:2019oyz} that uses the AMPT multiphase transport model~\cite{Lin:2004en}, coupled with a coalescence afterburner with Gaussian Wigner function to generate \fzero in three configurations, i.e., as a $\rm{s\bar{s}}$ meson, as a $\rm{u\bar{u}s\bar{s}}$ tetraquark state, and as a $\rm{K^+K^-}$ molecule.
The AMPT model contains four main components namely initial conditions, partonic interactions, conversion from partonic to hadronic matter, and interactions among hadrons based on a relativistic transport (ART) model~\cite{Li:1995pra,Li:2001xh}. The initial conditions are obtained from the HIJING model~\cite{Wang:1991hta} and the partonic interactions are determined according to the Zhang's Parton Cascade model~\cite{Zhang:1997ej}. 
In~\cite{Gu:2019oyz}, the authors use the phase-space information of quarks from this stage to implement quark coalescence for the \fzero with the $\rm{s\bar{s}}$ and tetraquark configurations.
In the default version of AMPT, the conversion of partons to hadrons is then calculated with the Lund string fragmentation~\cite{Andersson:1983jt,Andersson:1983ia,Sjostrand:1993yb}, while in the string melting version of the model~\cite{Lin:2001zk}, a quark coalescence approach is used to combine partons to form hadrons. 
The phase-space information of kaons generated at this stage by AMPT is used as input for the coalescence afterburner for the \fzero molecular state.
As shown in Fig.~\ref{fig:ptspectrum}, the \ssbar
calculation underestimates the \fzero \pt distribution by a factor of about three, whereas the molecule and the tetraquark configuration predictions are two and three orders of magnitude lower, respectively. Note that the molecule and the tetraquark configuration prediction ratios to data are reported in the lowest panel of Fig.~\ref{fig:ptspectrum}~ multiplied by a factor of 100 to improve the visibility. 
In addition, the shape of the \pt spectra for the $\rm{s\bar{s}}$ and the $\rm{u\bar{u}s\bar{s}}$ tetraquark configurations are found to be significantly steeper than the measured one. Instead, the ratio between the model prediction for the $\rm{K^+K^-}$ molecule configuration and the data exhibits a milder \pt dependence within uncertainties in the considered \pt range (0--3.5 \GeVc), indicating that in this configuration the model can reproduce qualitatively better the measured spectral shape.
Recent theoretical calculations that investigate the inclusive \fzero production according to the colour-singlet gluon-gluon fusion and colour evaporation model have been proposed by the authors of~\cite{Lebiedowicz:2020bwo}. As these exploratory studies are currently available only for \seven, a comparison with the data presented in this letter is auspicable in the near future.

The per-event \pt-integrated yield, d$N$/d$y$, and average transverse momentum, $\langle p_{\mathrm{T}} \rangle$, are calculated by integrating the \pt-differential yield in the measured transverse momentum range. The obtained values are the following:
\begin{eqnarray}
    \frac{{\rm{d}}N}{{\rm{d}}y} = 0.0385 \pm 0.0001 (\rm{stat.})  \pm~0.0047 (\rm{syst.}) \\
     \langle p_{\mathrm{T}} \rangle =  0.9624 \pm 0.0014 (\rm{stat.}) \pm~0.0357 (\rm{syst.}) \:\rm{Ge}\kern-.1em\rm{V}/c\xspace
\end{eqnarray}
 
Notably, the yield for \pt$>16$ \GeVc has a negligible contribution to the d$N$/d$y$ and thus no extrapolation was employed.
\begin{figure}[t]
\begin{center}
\includegraphics[scale=0.275]{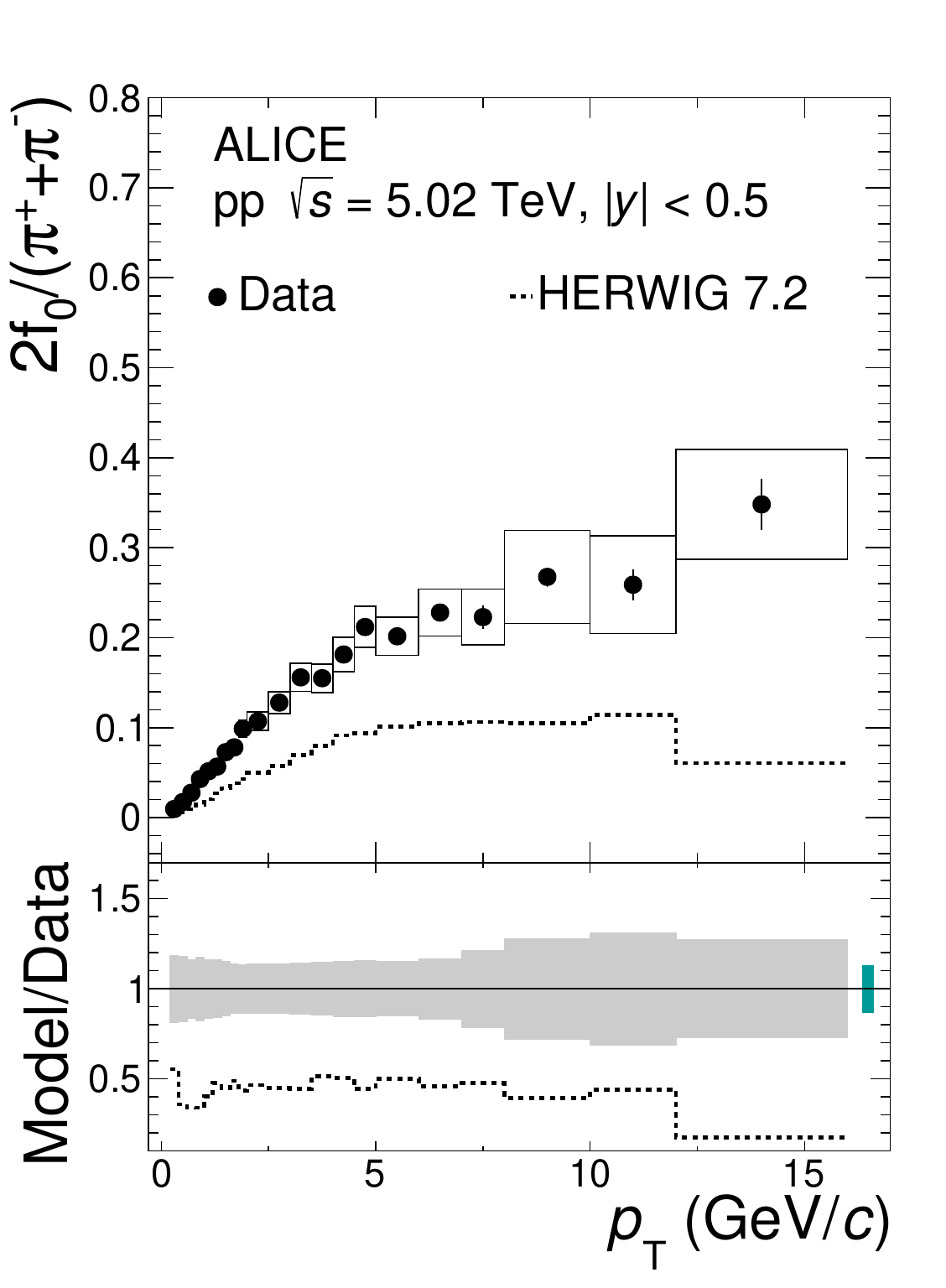}
\includegraphics[scale=0.275]{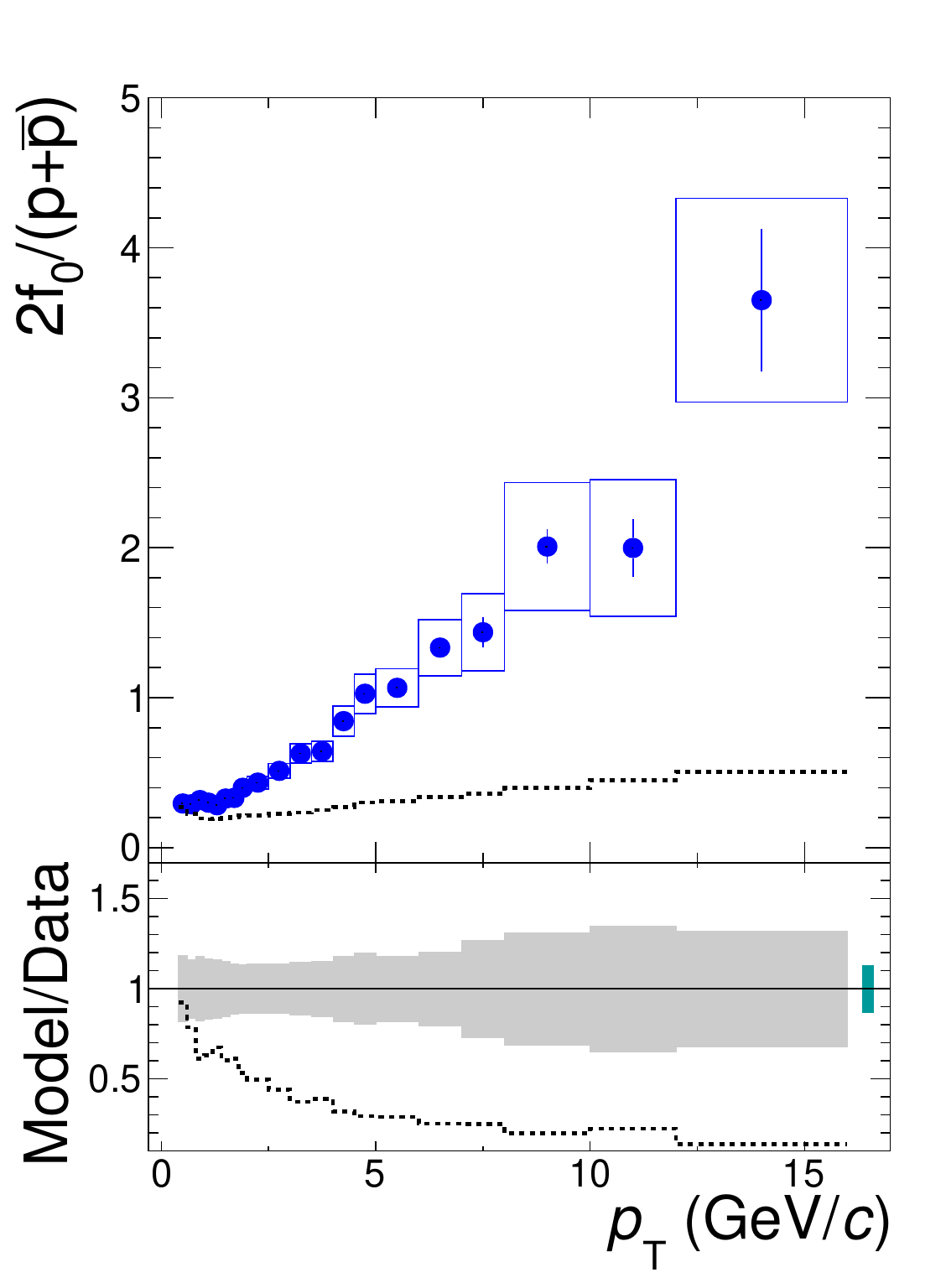}
\includegraphics[scale=0.275]{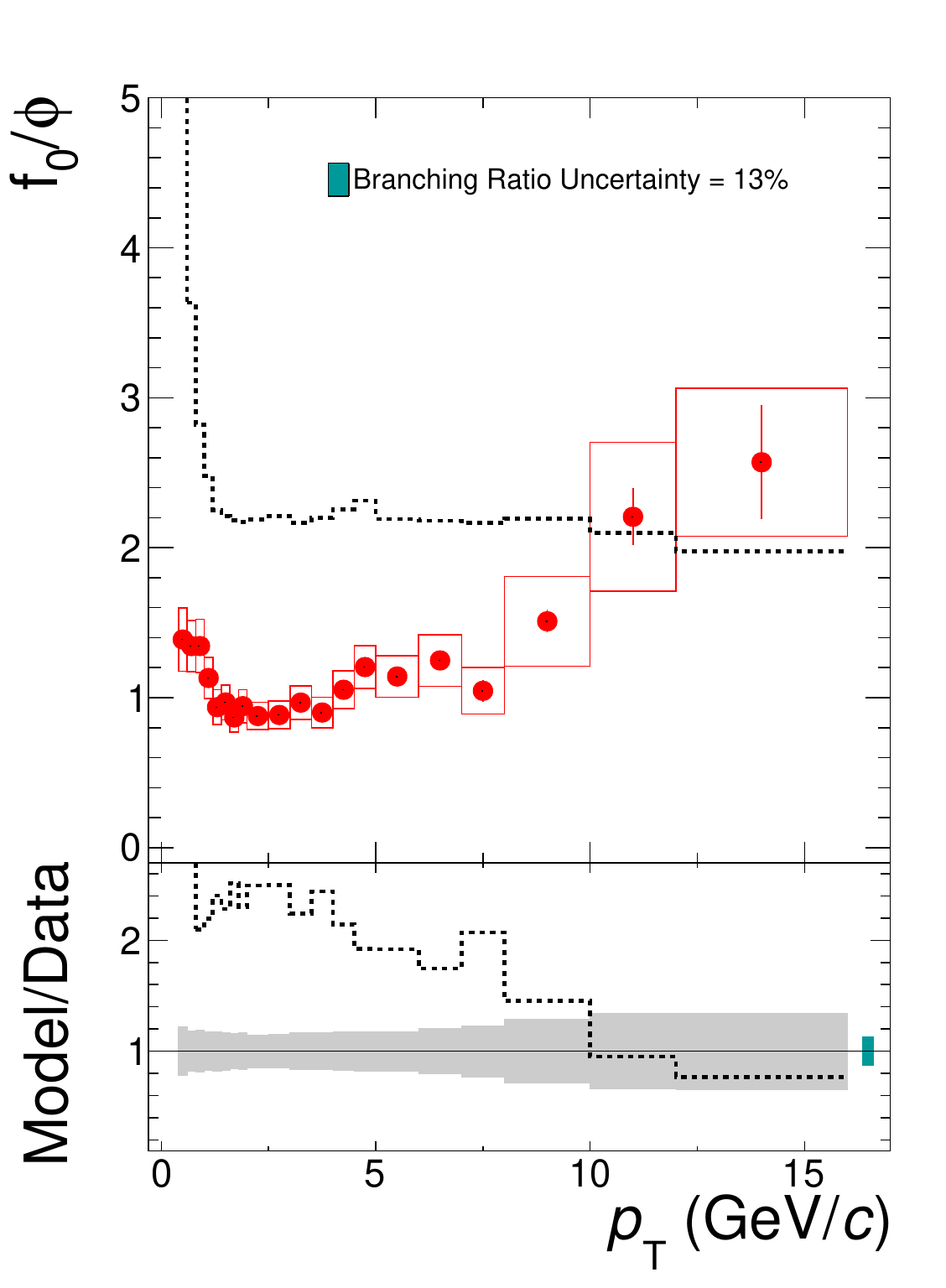}
\caption{(Upper panels) Particle yield ratios of \fzero to $\pi^{+}+\pi^{-}$~\cite{ALICE:2019hno} (left panel), $\rm{p}+\overline{\rm{p}}$~\cite{ALICE:2019hno} (middle panel), and $\phi$~\cite{ALICE:2019xyr} (right panel) measured in inelastic pp collisions at $\sqrt{s}=$ 5.02 TeV as a function of $p_{\mathrm{T}}$. Data are compared to HERWIG 7.2 model predictions. The statistical and systematic uncertainties are shown as bars and boxes, respectively.
(Lower panels) Ratio of measured particle ratios to the HERWIG model calculations (dashed histogram). The grey boxes at unity represent the sum in quadrature of the statistical and systematic uncertainty on the data. In the right panel, the ratio in the region for $\pt<0.8$~\GeVc is off-scale. The relative uncertainty of 13\% due to the branching ratio correction~\cite{Stone:2013eaa} is shown as a green box with an arbitrary horizontal width for visibility.}\label{fig:ptdiffratios}
\end{center}
\end{figure}

The production of \fzero is compared to that of other light-flavour hadrons in Fig.~\ref{fig:ptdiffratios} where the ratios of the \fzero yield to those of \pip+\pim~\cite{ALICE:2019hno}, 
p+\pbar~\cite{ALICE:2019hno}, 
and $\phi$~\cite{ALICE:2019xyr} measured in pp collisions at \five are reported as a function of \pt.
The ratio to \pip+\pim mesons exhibits an increasing trend as a function of $p_{\rm T}$ at low \pt and for $p_{\rm T}>$ 5 \GeVc it saturates within uncertainties. 

The comparison of the production of \fzero to that of protons and of the $\phi$ meson is particularly interesting as these particles have similar masses~\cite{Zyla:2020zbs} but different quark content. In particular, the $\phi$ meson is a pure \ssbar state, while the \fzero contains a light flavour component (\uubar, \ddbar) as well as a large \ssbar component, as suggested by measurements of \fzero produced in D$^{+}_{\rm{s}}$ decays~\cite{E791:2000lzz}. 
 The \fzero to $\rm{p}+\overline{\rm{p}}$ ratio shows an increasing monotonic trend as a function of $p_{\rm T}$, whereas the \fzero to $\phi$ ratio decreases for $p_{\rm T}<$ 1.5 \GeVc, remains flat till $p_{\rm T}\simeq$ 8 \GeVc, and increases for $p_{\rm T}>$ 8 \GeVc.

The measured \pt-differential particle yield ratios are compared in Fig.~\ref{fig:ptdiffratios} to the predictions from the HERWIG 7.2 event generator. 
The shape of the measured 2\fzero/(\pip+\pim) ratio is fairly well reproduced over almost the entire measured \pt range, although the yield is underestimated by about a factor of two by the model. 
The model underestimates the 2\fzero/(p+\pbar) ratio and fails to reproduce its \pt-dependence, with the measured ratio being more steeply increasing with \pt than the predicted one.
The trend of \fzero/$\phi$ ratio is flat for $1< \pt < 10$~\GeVc suggesting that its \pt dependence is qualitatively well reproduced by HERWIG in this momentum interval. However, the model overestimates the ratio by nearly a factor of two. For $\pt< 1$~\GeVc, the \fzero/$\phi$ ratio exhibits a steeply decreasing trend, that is qualitatively present also in the model prediction. At high \pt, the HERWIG predictions are consistent with the \fzero/$\phi$ data within the uncertainties. 

\begin{figure}[t]
\begin{center}
\includegraphics[width=0.6\linewidth]{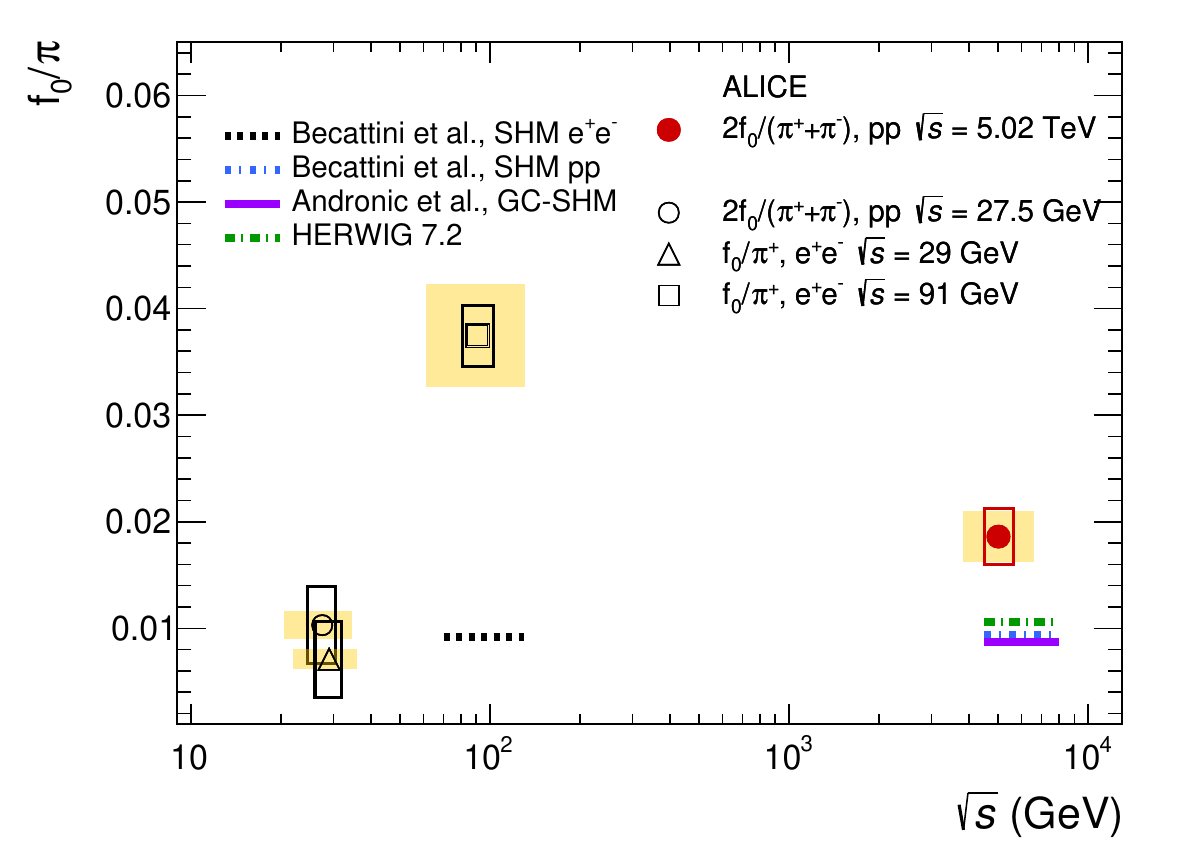}
\caption{Comparison of the measured $2\fzeroshort/(\pi^{+}+\pi^{-})$ ratio with measurements in \ee collisions at $\sqrt{s}$ = 29 \GeV~\cite{Hofmann:1988gy}, $\sqrt{s}$ = 91 \GeV~\cite{Chliapnikov:1999qi} and in pp collisions at $\sqrt{s}$ = 27.5 \GeV~\cite{Aguilar-Benitez:1991hzq}. The ratios are compared to predictions from statistical hadronisation model (SHM) calculations for \ee collisions~\cite{Becattini:2008tx} and pp collisions~\cite{Becattini:1997rv}, GC-SHM~\cite{Andronic:2017pug} and HERWIG 7.2~\cite{Bellm:2015jjp,Bahr:2008pv}. The hollow boxes represent the total uncertainty on data. The relative uncertainty of 13\% due to the branching ratio correction~\cite{Stone:2013eaa} applies to all data points and is shown as a yellow box.
All error boxes are drawn with an arbitrary horizontal width for visibility.}
\label{fig:ratios_vs_E}
\end{center}
\end{figure}

The ratio of the \pt-integrated \fzero yield relative to pions in \pp collisions at \five amounts to $2\fzeroshort/(\pip+\pim) = (0.0186 \pm 0.0026)$, with the uncertainty being the sum in quadrature of the statistical and systematic uncertainties. 
The value is shown in Fig.~\ref{fig:ratios_vs_E} (red point) in comparison with results from measurements in \pp and \ee collisions at lower centre-of-mass energies as well as with model calculations.

% table with the ratio values for each point for the references
\begin{figure}
\begin{center}
\includegraphics[width=0.6\linewidth]{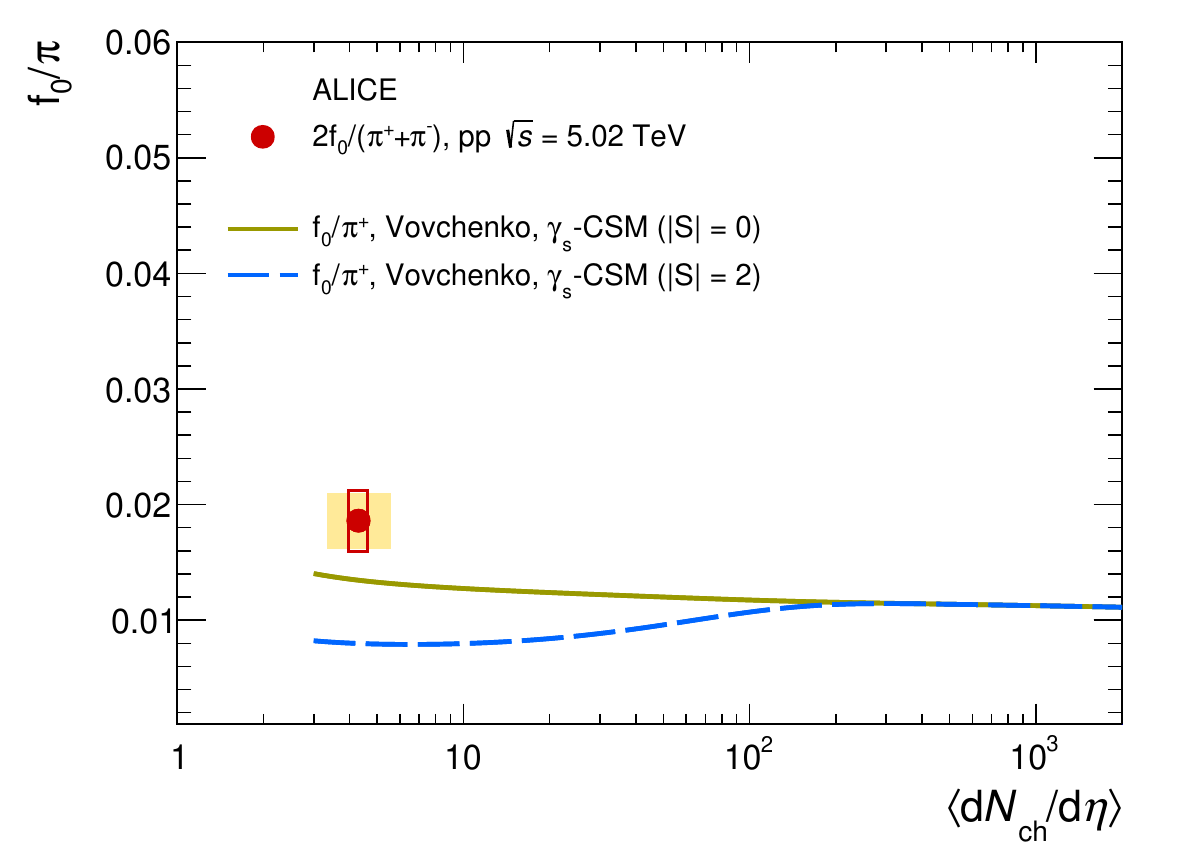}
\caption{
The $2\fzeroshort/(\pi^{+}+\pi^{-})$ ratio measured in pp collisions at \five~\cite{ALICE:2019hno} is compared to two distinct predictions for $\fzeroshort/\pi^{+}$ from a canonical statistical model ($\gamma_{\mathrm{s}}-$CSM~\cite{Vovchenko:2019kes}, see text for details) as a function of \avdndeta. The two calculations differ by the assumed strangeness content of \fzero and correspond to zero total strangeness, \mbox{$|S| = 0$} and \mbox{$|S| = 2$}. 
The height of the hollow red box represents the total uncertainty on the ratio, its width represents the uncertainty on the \avdndeta. The relative uncertainty of 13\% due to the branching ratio correction~\cite{Stone:2013eaa} is shown as a yellow box with an arbitrary horizontal width for visibility.}
\label{fig:ratios_vs_dndeta}
\end{center}
\end{figure}
The low energy experiment results were originally reported using different branching ratios, therefore all of them have been updated to take into account the most recent value of 46\%~\cite{Stone:2013eaa} used in this letter.  In Fig.~\ref{fig:ratios_vs_E}, the same uncertainty on the BR is applied to all data points and reported as a shaded yellow box.
The particle ratio value from the  fixed-target NA27 experiment at the CERN SPS, measured in \pp collisions at \s = 27.5~GeV~\cite{Aguilar-Benitez:1991hzq} is 44.5\% lower than the ratio measured at \five , suggesting a mild increase of the \fzero yield relative to pions with increasing energy of the pp collisions. The particle ratio values from \ee collisions at \s = 29 GeV~\cite{Hofmann:1988gy} and \s = 91 GeV~\cite{Chliapnikov:1999qi} are lower by 61\% and higher by a factor of two, respectively.
The particle ratios are compared with predictions based on  statistical hadronisation models~\cite{Becattini:2008tx,Becattini:1997rv,Andronic:2017pug} and the HERWIG 7.2 event generator. The statistical hadronisation model predictions by Becattini et al. for the \ee case~\cite{Becattini:2008tx} and for the \pp collisions~\cite{Becattini:1997rv} case underestimate the measurement by about a factor of two, similarly to a Grand Canonical formulation of the statistical hadronisation model (GC-SHM) from the GSI-Heidelberg group~\cite{Andronic:2017pug}. The value from HERWIG is also a factor of about two lower than the measured ratio.

In Fig.~\ref{fig:ratios_vs_dndeta}, the measured \pt-integrated $2\fzeroshort/(\pip+\pim)$ ratio in pp collisions at \five is compared to predictions based on the canonical statistical hadronisation model (CSM) described in~\cite{Vovchenko:2019kes}, as a function of the multiplicity of particles produced in the collision, expressed in terms of the average pseudorapidity density of charged particles, $\langle\rm{d}\it{N}_{ch}/\rm{d}\it{\eta}\rangle$. The prediction spans a large $\langle\rm{d}\it{N}_{ch}/\rm{d}\it{\eta}\rangle$ interval, reaching the high multiplicity achieved in central heavy-ion collisions at LHC energies. 
In canonical statistical hadronisation models, hadrons are formed from a source that is assumed to have reached full chemical equilibrium at the chemical freeze-out temperature $T_{\mathrm{ch}}$, and their yields are determined from the partition function for a canonical ensemble. The multiplicity dependence of hadron production is driven by the canonical suppression, namely the exact conservation of baryon number, electric charge, and strangeness over the correlation volume. 
The model considered here, with a temperature $T_{\mathrm{ch}} = 155$~MeV and a correlation volume that spans three units of rapidity, is able to reproduce the multiplicity dependence of hadron-to-pion ratios of several species over the charged particle multiplicity range covered by the ALICE measurements at the LHC, both qualitatively and quantitatively in most cases (see Fig.~5 of~\cite{Vovchenko:2019kes}).
In addition, to describe the multiplicity dependence of the $\phi/\pi$ ratio observed at the LHC, the model, henceforth labelled as $\gamma_{\mathrm{s}}-\mathrm{CSM}$, incorporates the incomplete equilibration of strangeness by introducing a strangeness saturation factor $\gamma_{\mathrm{s}}\leq 1$. Notably, in the strangeness nonequilibrium picture a \ssbar pair like the $\phi$ meson is effectively a double-strange particle ($\vert$S$\vert$ = 2), and ALICE $\phi$ data seem to be best described with $\vert$S$\vert$ = 1--2~\cite{ALICE:2014jbq}. 
The \fzero/\pip ratio is calculated for two scenarios, first assuming that the total strangeness content of \fzero is equal to zero (yellow continuous line) and second, assuming a total strangeness content equal to two (blue dashed line).
At high multiplicity, where the strangeness content of the system saturates in presence of a QGP, the calculations for the two scenarios converge and reach the grand canonical limit.

In the first scenario ($|S|$ = 0), $\gamma_{\mathrm{s}}-$CSM predicts higher values for the \fzero to pion yield ratio as compared to the second scenario ($|S|$ = 2) in the low $\langle \rm{d}\it{N}_{\rm{ch}}/\rm{d}\it{\eta}\rangle$ region. The two predictions match each other for $\langle \rm{d}\it{N}_{\rm{ch}}/\rm{d}\it{\eta}\rangle$ $\geq$ 100. 
The measured $2\fzeroshort/(\pi^{+}+\pi^{-})$ ratio in pp collisions at \five differs by 1.9$\sigma$ from the $\gamma_{\mathrm{s}}-$CSM prediction for the \fzero with net strangeness equal to zero, and by 4.0$\sigma$ from the $|S|$ = 2 prediction, indicating that the former scenario is favoured over the latter in this implementation of the model.

%% file: conclusions.tex
\section{Conclusions} 
\label{sec:Conclusions}
In conclusion, the first measurement of \fzero production in inelastic \pp collisions at \five at the LHC is presented. The measurement is performed in a wide \pt interval from 0 to 16 \GeVc at midrapidity by reconstructing the resonance in the hadronic decay channel \mbox{\fzero$\rightarrow\pip\pim$}. 
The inclusive \fzero production is underestimated by HERWIG 7.2 by a factor of two for $1<\pt<4$~\GeVc and by a large factor (up to  more than four) in $4<\pt<16$~\GeVc. However, this QCD-inspired event generator is able to describe the \pt-dependence of the 2\fzero/(\pip+\pim) and \fzero/$\phi$ ratios in a rather broad \pt range, while failing to reproduce the 2\fzero/(p+\pbar) ratio over the entire measured \pt range. 
The production of \fzero is also not described by a AMPT+coalescence model prediction in three configurations ($\rm{s\bar{s}}$ meson, $\rm{u\bar{u}s\bar{s}}$ tetraquark state, and $\rm{K^+K^-}$ molecule), which use the phase space information of quarks and kaons from the AMPT model. 
In order to compare the new measurement of the \pt-integrated \fzero to pion ratio with low energy data, the low energy points were updated with the latest branching ratio. The new result presented in this letter suggests a mild increase of the production of \fzero relative to pions in inelastic \pp collisions from \mbox{$\sqrt{s}$ = 27.5\,\GeV} to \five.   
For the same ratio, \mbox{HERWIG 7.2} predicts a value that is about 43\% lower than the measured one whereas different implementations of the statistical hadronisation model underestimate the data by up to a factor of about two. Notably, the $\gamma_{s}-$CSM prediction for the \fzero assuming net strangeness equal to zero is consistent with the data within 1.9$\sigma$.  
 
In summary, the description of the inclusive \fzeroshort production in \pp collisions provided by the few event generators and theoretical calculations that attempt its modelling is, at present, unsatisfactory. Future developments in this direction may help gaining insight over the nature of this particle, as new data may become available.
From the experimental point of view, the results presented in this letter set the necessary baseline for the future measurements of the production and the nuclear modification factor of \fzero in \pPb and \PbPb collisions at the LHC, which have been suggested as observables that are sensitive to the elusive nature of this particle.

%% file: fa_2022-06-03_Opt_C.tex
% Version: 2022-06-03

The ALICE Collaboration would like to thank all its engineers and technicians for their invaluable contributions to the construction of the experiment and the CERN accelerator teams for the outstanding performance of the LHC complex.
The ALICE Collaboration gratefully acknowledges the resources and support provided by all Grid centres and the Worldwide LHC Computing Grid (WLCG) collaboration.
The ALICE Collaboration acknowledges the following funding agencies for their support in building and running the ALICE detector:
A. I. Alikhanyan National Science Laboratory (Yerevan Physics Institute) Foundation (ANSL), State Committee of Science and World Federation of Scientists (WFS), Armenia;
Austrian Academy of Sciences, Austrian Science Fund (FWF): [M 2467-N36] and Nationalstiftung f\"{u}r Forschung, Technologie und Entwicklung, Austria;
Ministry of Communications and High Technologies, National Nuclear Research Center, Azerbaijan;
Conselho Nacional de Desenvolvimento Cient\'{\i}fico e Tecnol\'{o}gico (CNPq), Financiadora de Estudos e Projetos (Finep), Funda\c{c}\~{a}o de Amparo \`{a} Pesquisa do Estado de S\~{a}o Paulo (FAPESP) and Universidade Federal do Rio Grande do Sul (UFRGS), Brazil;
Bulgarian Ministry of Education and Science, within the National Roadmap for Research Infrastructures 2020-2027 (object CERN), Bulgaria;
Ministry of Education of China (MOEC) , Ministry of Science \& Technology of China (MSTC) and National Natural Science Foundation of China (NSFC), China;
Ministry of Science and Education and Croatian Science Foundation, Croatia;
Centro de Aplicaciones Tecnol\'{o}gicas y Desarrollo Nuclear (CEADEN), Cubaenerg\'{\i}a, Cuba;
Ministry of Education, Youth and Sports of the Czech Republic, Czech Republic;
The Danish Council for Independent Research | Natural Sciences, the VILLUM FONDEN and Danish National Research Foundation (DNRF), Denmark;
Helsinki Institute of Physics (HIP), Finland;
Commissariat \`{a} l'Energie Atomique (CEA) and Institut National de Physique Nucl\'{e}aire et de Physique des Particules (IN2P3) and Centre National de la Recherche Scientifique (CNRS), France;
Bundesministerium f\"{u}r Bildung und Forschung (BMBF) and GSI Helmholtzzentrum f\"{u}r Schwerionenforschung GmbH, Germany;
General Secretariat for Research and Technology, Ministry of Education, Research and Religions, Greece;
National Research, Development and Innovation Office, Hungary;
Department of Atomic Energy Government of India (DAE), Department of Science and Technology, Government of India (DST), University Grants Commission, Government of India (UGC) and Council of Scientific and Industrial Research (CSIR), India;
National Research and Innovation Agency - BRIN, Indonesia;
Istituto Nazionale di Fisica Nucleare (INFN), Italy;
Japanese Ministry of Education, Culture, Sports, Science and Technology (MEXT) and Japan Society for the Promotion of Science (JSPS) KAKENHI, Japan;
Consejo Nacional de Ciencia (CONACYT) y Tecnolog\'{i}a, through Fondo de Cooperaci\'{o}n Internacional en Ciencia y Tecnolog\'{i}a (FONCICYT) and Direcci\'{o}n General de Asuntos del Personal Academico (DGAPA), Mexico;
Nederlandse Organisatie voor Wetenschappelijk Onderzoek (NWO), Netherlands;
The Research Council of Norway, Norway;
Commission on Science and Technology for Sustainable Development in the South (COMSATS), Pakistan;
Pontificia Universidad Cat\'{o}lica del Per\'{u}, Peru;
Ministry of Education and Science, National Science Centre and WUT ID-UB, Poland;
Korea Institute of Science and Technology Information and National Research Foundation of Korea (NRF), Republic of Korea;
Ministry of Education and Scientific Research, Institute of Atomic Physics, Ministry of Research and Innovation and Institute of Atomic Physics and University Politehnica of Bucharest, Romania;
Ministry of Education, Science, Research and Sport of the Slovak Republic, Slovakia;
National Research Foundation of South Africa, South Africa;
Swedish Research Council (VR) and Knut \& Alice Wallenberg Foundation (KAW), Sweden;
European Organization for Nuclear Research, Switzerland;
Suranaree University of Technology (SUT), National Science and Technology Development Agency (NSTDA) and National Science, Research and Innovation Fund (NSRF via PMU-B B05F650021), Thailand;
Turkish Energy, Nuclear and Mineral Research Agency (TENMAK), Turkey;
National Academy of  Sciences of Ukraine, Ukraine;
Science and Technology Facilities Council (STFC), United Kingdom;
National Science Foundation of the United States of America (NSF) and United States Department of Energy, Office of Nuclear Physics (DOE NP), United States of America.
In addition, individual groups or members have received support from:
Marie Sk\l{}odowska Curie, European Research Council, Strong 2020 - Horizon 2020 (grant nos. 950692, 824093, 896850), European Union;
Academy of Finland (Center of Excellence in Quark Matter) (grant nos. 346327, 346328), Finland;
Programa de Apoyos para la Superaci\'{o}n del Personal Acad\'{e}mico, UNAM, Mexico.

%% file: 2022-06-03-Alice_Authorlist_2022-06-03_0_Opt_C.tex
% ALICE Collaboration author list for 2022-06-03
\begin{flushleft} 
\small

S.~Acharya\,\orcidlink{0000-0002-9213-5329}\,$^{\rm 124}$, 
D.~Adamov\'{a}\,\orcidlink{0000-0002-0504-7428}\,$^{\rm 85}$, 
A.~Adler$^{\rm 69}$, 
G.~Aglieri Rinella\,\orcidlink{0000-0002-9611-3696}\,$^{\rm 32}$, 
M.~Agnello\,\orcidlink{0000-0002-0760-5075}\,$^{\rm 29}$, 
N.~Agrawal\,\orcidlink{0000-0003-0348-9836}\,$^{\rm 50}$, 
Z.~Ahammed\,\orcidlink{0000-0001-5241-7412}\,$^{\rm 132}$, 
S.~Ahmad\,\orcidlink{0000-0003-0497-5705}\,$^{\rm 15}$, 
S.U.~Ahn\,\orcidlink{0000-0001-8847-489X}\,$^{\rm 70}$, 
I.~Ahuja\,\orcidlink{0000-0002-4417-1392}\,$^{\rm 37}$
% ...

\section*{Affiliation Notes}

$^{\rm I}$ Deceased\\
$^{\rm II}$ Also at: Max-Planck-Institut f\"{u}r Physik, Munich, Germany\\
$^{\rm III}$ Also at: Italian National Agency for New Technologies, Energy and Sustainable Economic Development (ENEA), Bologna, Italy\\
$^{\rm IV}$ Also at: Dipartimento DET del Politecnico di Torino, Turin, Italy\\
$^{\rm V}$ Also at: Department of Applied Physics, Aligarh Muslim University, Aligarh, India\\
$^{\rm VI}$ Also at: Institute of Theoretical Physics, University of Wroclaw, Poland\\
$^{\rm VII}$ Also at: An institution covered by a cooperation agreement with CERN\\

\section*{Collaboration Institutes}

$^{1}$ A.I. Alikhanyan National Science Laboratory (Yerevan Physics Institute) Foundation, Yerevan, Armenia\\
$^{2}$ AGH University of Krakow, Cracow, Poland\\
$^{3}$ Bogolyubov Institute for Theoretical Physics, National Academy of Sciences of Ukraine, Kiev, Ukraine\\
$^{4}$ Bose Institute, Department of Physics  and Centre for Astroparticle Physics and Space Science (CAPSS), Kolkata, India\\
$^{5}$ California Polytechnic State University, San Luis Obispo, California, United States\\
$^{6}$ Central China Normal University, Wuhan, China\\
$^{7}$ Centro de Aplicaciones Tecnol\'{o}gicas y Desarrollo Nuclear (CEADEN), Havana, Cuba\\
$^{8}$ Centro de Investigaci\'{o}n y de Estudios Avanzados (CINVESTAV), Mexico City and M\'{e}rida, Mexico\\
$^{9}$ Chicago State University, Chicago, Illinois, United States\\
$^{10}$ China Institute of Atomic Energy, Beijing, China\\
$^{11}$ Chungbuk National University, Cheongju, Republic of Korea\\
$^{12}$ Comenius University Bratislava, Faculty of Mathematics, Physics and Informatics, Bratislava, Slovak Republic\\
$^{13}$ COMSATS University Islamabad, Islamabad, Pakistan\\
$^{14}$ Creighton University, Omaha, Nebraska, United States\\
$^{15}$ Department of Physics, Aligarh Muslim University, Aligarh, India\\
$^{16}$ Department of Physics, Pusan National University, Pusan, Republic of Korea\\
$^{17}$ Department of Physics, Sejong University, Seoul, Republic of Korea\\
$^{18}$ Department of Physics, University of California, Berkeley, California, United States\\
$^{19}$ Department of Physics, University of Oslo, Oslo, Norway\\
$^{20}$ Department of Physics and Technology, University of Bergen, Bergen, Norway\\
$^{21}$ Dipartimento di Fisica, Universit\`{a} di Pavia, Pavia, Italy\\
$^{22}$ Dipartimento di Fisica dell'Universit\`{a} and Sezione INFN, Cagliari, Italy\\
$^{23}$ Dipartimento di Fisica dell'Universit\`{a} and Sezione INFN, Trieste, Italy\\
$^{24}$ Dipartimento di Fisica dell'Universit\`{a} and Sezione INFN, Turin, Italy\\
$^{25}$ Dipartimento di Fisica e Astronomia dell'Universit\`{a} and Sezione INFN, Bologna, Italy\\
$^{26}$ Dipartimento di Fisica e Astronomia dell'Universit\`{a} and Sezione INFN, Catania, Italy\\
$^{27}$ Dipartimento di Fisica e Astronomia dell'Universit\`{a} and Sezione INFN, Padova, Italy\\
$^{28}$ Dipartimento di Fisica `E.R.~Caianiello' dell'Universit\`{a} and Gruppo Collegato INFN, Salerno, Italy\\
$^{29}$ Dipartimento DISAT del Politecnico and Sezione INFN, Turin, Italy\\
$^{30}$ Dipartimento di Scienze MIFT, Universit\`{a} di Messina, Messina, Italy\\
$^{31}$ Dipartimento Interateneo di Fisica `M.~Merlin' and Sezione INFN, Bari, Italy\\
$^{32}$ European Organization for Nuclear Research (CERN), Geneva, Switzerland\\
$^{33}$ Faculty of Electrical Engineering, Mechanical Engineering and Naval Architecture, University of Split, Split, Croatia\\
$^{34}$ Faculty of Engineering and Science, Western Norway University of Applied Sciences, Bergen, Norway\\
$^{35}$ Faculty of Nuclear Sciences and Physical Engineering, Czech Technical University in Prague, Prague, Czech Republic\\
$^{36}$ Faculty of Physics, Sofia University, Sofia, Bulgaria\\
$^{37}$ Faculty of Science, P.J.~\v{S}af\'{a}rik University, Ko\v{s}ice, Slovak Republic\\
$^{38}$ Frankfurt Institute for Advanced Studies, Johann Wolfgang Goethe-Universit\"{a}t Frankfurt, Frankfurt, Germany\\
$^{39}$ Fudan University, Shanghai, China\\
$^{40}$ Gangneung-Wonju National University, Gangneung, Republic of Korea\\
$^{41}$ Gauhati University, Department of Physics, Guwahati, India\\
$^{42}$ Helmholtz-Institut f\"{u}r Strahlen- und Kernphysik, Rheinische Friedrich-Wilhelms-Universit\"{a}t Bonn, Bonn, Germany\\
$^{43}$ Helsinki Institute of Physics (HIP), Helsinki, Finland\\
$^{44}$ High Energy Physics Group,  Universidad Aut\'{o}noma de Puebla, Puebla, Mexico\\
$^{45}$ Horia Hulubei National Institute of Physics and Nuclear Engineering, Bucharest, Romania\\
$^{46}$ Indian Institute of Technology Bombay (IIT), Mumbai, India\\
$^{47}$ Indian Institute of Technology Indore, Indore, India\\
$^{48}$ INFN, Laboratori Nazionali di Frascati, Frascati, Italy\\
$^{49}$ INFN, Sezione di Bari, Bari, Italy\\
$^{50}$ INFN, Sezione di Bologna, Bologna, Italy\\
$^{51}$ INFN, Sezione di Cagliari, Cagliari, Italy\\
$^{52}$ INFN, Sezione di Catania, Catania, Italy\\
$^{53}$ INFN, Sezione di Padova, Padova, Italy\\
$^{54}$ INFN, Sezione di Pavia, Pavia, Italy\\
$^{55}$ INFN, Sezione di Torino, Turin, Italy\\
$^{56}$ INFN, Sezione di Trieste, Trieste, Italy\\
$^{57}$ Inha University, Incheon, Republic of Korea\\
$^{58}$ Institute for Gravitational and Subatomic Physics (GRASP), Utrecht University/Nikhef, Utrecht, Netherlands\\
$^{59}$ Institute of Experimental Physics, Slovak Academy of Sciences, Ko\v{s}ice, Slovak Republic\\
$^{60}$ Institute of Physics, Homi Bhabha National Institute, Bhubaneswar, India\\
$^{61}$ Institute of Physics of the Czech Academy of Sciences, Prague, Czech Republic\\
$^{62}$ Institute of Space Science (ISS), Bucharest, Romania\\
$^{63}$ Institut f\"{u}r Kernphysik, Johann Wolfgang Goethe-Universit\"{a}t Frankfurt, Frankfurt, Germany\\
$^{64}$ Instituto de Ciencias Nucleares, Universidad Nacional Aut\'{o}noma de M\'{e}xico, Mexico City, Mexico\\
$^{65}$ Instituto de F\'{i}sica, Universidade Federal do Rio Grande do Sul (UFRGS), Porto Alegre, Brazil\\
$^{66}$ Instituto de F\'{\i}sica, Universidad Nacional Aut\'{o}noma de M\'{e}xico, Mexico City, Mexico\\
$^{67}$ iThemba LABS, National Research Foundation, Somerset West, South Africa\\
$^{68}$ Jeonbuk National University, Jeonju, Republic of Korea\\
$^{69}$ Johann-Wolfgang-Goethe Universit\"{a}t Frankfurt Institut f\"{u}r Informatik, Fachbereich Informatik und Mathematik, Frankfurt, Germany\\
$^{70}$ Korea Institute of Science and Technology Information, Daejeon, Republic of Korea\\
$^{71}$ KTO Karatay University, Konya, Turkey\\
$^{72}$ Laboratoire de Physique Subatomique et de Cosmologie, Universit\'{e} Grenoble-Alpes, CNRS-IN2P3, Grenoble, France\\
$^{73}$ Lawrence Berkeley National Laboratory, Berkeley, California, United States\\
$^{74}$ Lund University Department of Physics, Division of Particle Physics, Lund, Sweden\\
$^{75}$ Nagasaki Institute of Applied Science, Nagasaki, Japan\\
$^{76}$ Nara Women{'}s University (NWU), Nara, Japan\\
$^{77}$ National and Kapodistrian University of Athens, School of Science, Department of Physics , Athens, Greece\\
$^{78}$ National Centre for Nuclear Research, Warsaw, Poland\\
$^{79}$ National Institute of Science Education and Research, Homi Bhabha National Institute, Jatni, India\\
$^{80}$ National Nuclear Research Center, Baku, Azerbaijan\\
$^{81}$ National Research and Innovation Agency - BRIN, Jakarta, Indonesia\\
$^{82}$ Niels Bohr Institute, University of Copenhagen, Copenhagen, Denmark\\
$^{83}$ Nikhef, National institute for subatomic physics, Amsterdam, Netherlands\\
$^{84}$ Nuclear Physics Group, STFC Daresbury Laboratory, Daresbury, United Kingdom\\
$^{85}$ Nuclear Physics Institute of the Czech Academy of Sciences, Husinec-\v{R}e\v{z}, Czech Republic\\
$^{86}$ Oak Ridge National Laboratory, Oak Ridge, Tennessee, United States\\
$^{87}$ Ohio State University, Columbus, Ohio, United States\\
$^{88}$ Physics department, Faculty of science, University of Zagreb, Zagreb, Croatia\\
$^{89}$ Physics Department, Panjab University, Chandigarh, India\\
$^{90}$ Physics Department, University of Jammu, Jammu, India\\
$^{91}$ Physics Department, University of Rajasthan, Jaipur, India\\
$^{92}$ Physics Program and International Institute for Sustainability with Knotted Chiral Meta Matter (SKCM2), Hiroshima University, Hiroshima, Japan\\
$^{93}$ Physikalisches Institut, Eberhard-Karls-Universit\"{a}t T\"{u}bingen, T\"{u}bingen, Germany\\
$^{94}$ Physikalisches Institut, Ruprecht-Karls-Universit\"{a}t Heidelberg, Heidelberg, Germany\\
$^{95}$ Physik Department, Technische Universit\"{a}t M\"{u}nchen, Munich, Germany\\
$^{96}$ Politecnico di Bari and Sezione INFN, Bari, Italy\\
$^{97}$ Research Division and ExtreMe Matter Institute EMMI, GSI Helmholtzzentrum f\"ur Schwerionenforschung GmbH, Darmstadt, Germany\\
$^{98}$ Saga University, Saga, Japan\\
$^{99}$ Saha Institute of Nuclear Physics, Homi Bhabha National Institute, Kolkata, India\\
$^{100}$ School of Physics and Astronomy, University of Birmingham, Birmingham, United Kingdom\\
$^{101}$ Secci\'{o}n F\'{\i}sica, Departamento de Ciencias, Pontificia Universidad Cat\'{o}lica del Per\'{u}, Lima, Peru\\
$^{102}$ Stefan Meyer Institut f\"{u}r Subatomare Physik (SMI), Vienna, Austria\\
$^{103}$ SUBATECH, IMT Atlantique, Nantes Universit\'{e}, CNRS-IN2P3, Nantes, France\\
$^{104}$ Suranaree University of Technology, Nakhon Ratchasima, Thailand\\
$^{105}$ Technical University of Ko\v{s}ice, Ko\v{s}ice, Slovak Republic\\
$^{106}$ The Henryk Niewodniczanski Institute of Nuclear Physics, Polish Academy of Sciences, Cracow, Poland\\
$^{107}$ The University of Texas at Austin, Austin, Texas, United States\\
$^{108}$ Universidad Aut\'{o}noma de Sinaloa, Culiac\'{a}n, Mexico\\
$^{109}$ Universidade de S\~{a}o Paulo (USP), S\~{a}o Paulo, Brazil\\
$^{110}$ Universidade Estadual de Campinas (UNICAMP), Campinas, Brazil\\
$^{111}$ Universidade Federal do ABC, Santo Andre, Brazil\\
$^{112}$ University of Cape Town, Cape Town, South Africa\\
$^{113}$ University of Houston, Houston, Texas, United States\\
$^{114}$ University of Jyv\"{a}skyl\"{a}, Jyv\"{a}skyl\"{a}, Finland\\
$^{115}$ University of Kansas, Lawrence, Kansas, United States\\
$^{116}$ University of Liverpool, Liverpool, United Kingdom\\
$^{117}$ University of Science and Technology of China, Hefei, China\\
$^{118}$ University of South-Eastern Norway, Kongsberg, Norway\\
$^{119}$ University of Tennessee, Knoxville, Tennessee, United States\\
$^{120}$ University of the Witwatersrand, Johannesburg, South Africa\\
$^{121}$ University of Tokyo, Tokyo, Japan\\
$^{122}$ University of Tsukuba, Tsukuba, Japan\\
$^{123}$ University Politehnica of Bucharest, Bucharest, Romania\\
$^{124}$ Universit\'{e} Clermont Auvergne, CNRS/IN2P3, LPC, Clermont-Ferrand, France\\
$^{125}$ Universit\'{e} de Lyon, CNRS/IN2P3, Institut de Physique des 2 Infinis de Lyon, Lyon, France\\
$^{126}$ Universit\'{e} de Strasbourg, CNRS, IPHC UMR 7178, F-67000 Strasbourg, France, Strasbourg, France\\
$^{127}$ Universit\'{e} Paris-Saclay, Centre d'Etudes de Saclay (CEA), IRFU, D\'{e}partment de Physique Nucl\'{e}aire (DPhN), Saclay, France\\
$^{128}$ Universit\'{e}  Paris-Saclay, CNRS/IN2P3, IJCLab, Orsay, France\\
$^{129}$ Universit\`{a} degli Studi di Foggia, Foggia, Italy\\
$^{130}$ Universit\`{a} del Piemonte Orientale, Vercelli, Italy\\
$^{131}$ Universit\`{a} di Brescia, Brescia, Italy\\
$^{132}$ Variable Energy Cyclotron Centre, Homi Bhabha National Institute, Kolkata, India\\
$^{133}$ Warsaw University of Technology, Warsaw, Poland\\
$^{134}$ Wayne State University, Detroit, Michigan, United States\\
$^{135}$ Westf\"{a}lische Wilhelms-Universit\"{a}t M\"{u}nster, Institut f\"{u}r Kernphysik, M\"{u}nster, Germany\\
$^{136}$ Wigner Research Centre for Physics, Budapest, Hungary\\
$^{137}$ Yale University, New Haven, Connecticut, United States\\
$^{138}$ Yonsei University, Seoul, Republic of Korea\\
$^{139}$  Zentrum  f\"{u}r Technologie und Transfer (ZTT), Worms, Germany\\
$^{140}$ Affiliated with an institute covered by a cooperation agreement with CERN\\
$^{141}$ Affiliated with an international laboratory covered by a cooperation agreement with CERN.\\

\end{flushleft} 

%% file: main.bbl
\providecommand{\href}[2]{#2}\begingroup\raggedright\begin{thebibliography}{10}

\bibitem{Gell-Mann:1964ewy}
M.~Gell-Mann, ``{A Schematic Model of Baryons and Mesons}'',
  \href{http://dx.doi.org/10.1016/S0031-9163(64)92001-3}{{\em Phys. Lett.}
  {\bfseries 8} (1964) 214--215}.

\bibitem{Zyla:2020zbs}
{\bfseries Particle Data Group} Collaboration, P.~Zyla {\em et~al.}, ``{Review
  of Particle Physics}'', \href{http://dx.doi.org/10.1093/ptep/ptaa104}{{\em
  PTEP} {\bfseries 2020} (2020) 083C01}.

\bibitem{Jaffe:1976ig}
R.~L. Jaffe, ``{Multi-Quark Hadrons. 1. The Phenomenology of (2 Quark 2
  anti-Quark) Mesons}'', \href{http://dx.doi.org/10.1103/PhysRevD.15.267}{{\em
  Phys. Rev. D} {\bfseries 15} (1977) 267}.

\bibitem{Jaffe:1976ih}
R.~L. Jaffe, ``{Multi-Quark Hadrons. 2. Methods}'',
  \href{http://dx.doi.org/10.1103/PhysRevD.15.281}{{\em Phys. Rev. D}
  {\bfseries 15} (1977) 281}.

\bibitem{Close:2002zu}
F.~E. Close and N.~A. Tornqvist, ``{Scalar mesons above and below 1-GeV}'',
  \href{http://dx.doi.org/10.1088/0954-3899/28/10/201}{{\em J. Phys. G}
  {\bfseries 28} (2002) R249--R267},
  \href{http://arxiv.org/abs/hep-ph/0204205}{{\ttfamily arXiv:hep-ph/0204205}}.

\bibitem{Amsler:2004ps}
C.~Amsler and N.~A. Tornqvist, ``{Mesons beyond the naive quark model}'',
  \href{http://dx.doi.org/10.1016/j.physrep.2003.09.003}{{\em Phys. Rept.}
  {\bfseries 389} (2004) 61--117}.

\bibitem{Maiani:2004uc}
L.~Maiani, F.~Piccinini, A.~D. Polosa, and V.~Riquer, ``{A New look at scalar
  mesons}'', \href{http://dx.doi.org/10.1103/PhysRevLett.93.212002}{{\em Phys.
  Rev. Lett.} {\bfseries 93} (2004) 212002},
  \href{http://arxiv.org/abs/hep-ph/0407017}{{\ttfamily arXiv:hep-ph/0407017}}.

\bibitem{Klempt:2007cp}
E.~Klempt and A.~Zaitsev, ``{Glueballs, Hybrids, Multiquarks. Experimental
  facts versus QCD inspired concepts}'',
  \href{http://dx.doi.org/10.1016/j.physrep.2007.07.006}{{\em Phys. Rept.}
  {\bfseries 454} (2007) 1--202},
  \href{http://arxiv.org/abs/0708.4016}{{\ttfamily arXiv:0708.4016 [hep-ph]}}.

\bibitem{Chen:2003za}
C.-H. Chen, ``{Evidence for two quark content of f(0)(980) in exclusive b
  ---\ensuremath{>} c decays}'',
  \href{http://dx.doi.org/10.1103/PhysRevD.67.094011}{{\em Phys. Rev. D}
  {\bfseries 67} (2003) 094011},
  \href{http://arxiv.org/abs/hep-ph/0302059}{{\ttfamily arXiv:hep-ph/0302059}}.

\bibitem{Achasov:2020aun}
N.~N. Achasov, J.~V. Bennett, A.~V. Kiselev, E.~A. Kozyrev, and G.~N.
  Shestakov, ``{Evidence of the four-quark nature of $f_0$(980) and
  $f_0$(500)}'', \href{http://dx.doi.org/10.1103/PhysRevD.103.014010}{{\em
  Phys. Rev. D} {\bfseries 103} (2021) 014010},
  \href{http://arxiv.org/abs/2009.04191}{{\ttfamily arXiv:2009.04191
  [hep-ph]}}.

\bibitem{Ahmed:2020kmp}
H.~A. Ahmed and C.~W. Xiao, ``{Study the molecular nature of $\sigma$,
  $f_{0}(980)$, and $a_{0}(980)$ states}'',
  \href{http://dx.doi.org/10.1103/PhysRevD.101.094034}{{\em Phys. Rev. D}
  {\bfseries 101} (2020) 094034},
  \href{http://arxiv.org/abs/2001.08141}{{\ttfamily arXiv:2001.08141
  [hep-ph]}}.

\bibitem{Oliinychenko:2021enj}
D.~Oliinychenko and C.~Shen, ``{Resonance production in PbPb collisions at 5.02
  TeV via hydrodynamics and hadronic afterburner}'',
  \href{http://arxiv.org/abs/2105.07539}{{\ttfamily arXiv:2105.07539
  [hep-ph]}}.

\bibitem{Achasov:2003cn}
N.~N. Achasov, ``{Radiative decays of phi meson about nature of light scalar
  resonances}'', \href{http://dx.doi.org/10.1016/j.nuclphysa.2003.09.002}{{\em
  Nucl. Phys. A} {\bfseries 728} (2003) 425--438},
  \href{http://arxiv.org/abs/hep-ph/0309118}{{\ttfamily arXiv:hep-ph/0309118}}.

\bibitem{Achasov:2000ku}
M.~N. Achasov {\em et~al.}, ``{The phi --\ensuremath{>} eta pi0 gamma decay}'',
  \href{http://dx.doi.org/10.1016/S0370-2693(00)00334-8}{{\em Phys. Lett. B}
  {\bfseries 479} (2000) 53--58},
  \href{http://arxiv.org/abs/hep-ex/0003031}{{\ttfamily arXiv:hep-ex/0003031}}.

\bibitem{CMD-2:1999imm}
{\bfseries CMD-2} Collaboration, R.~R. Akhmetshin {\em et~al.}, ``{First
  observation of the phi --\ensuremath{>} pi+ pi- gamma decay}'',
  \href{http://dx.doi.org/10.1016/S0370-2693(99)00919-3}{{\em Phys. Lett. B}
  {\bfseries 462} (1999) 371},
  \href{http://arxiv.org/abs/hep-ex/9907005}{{\ttfamily arXiv:hep-ex/9907005}}.

\bibitem{KLOE:2002kzf}
{\bfseries KLOE} Collaboration, A.~Aloisio {\em et~al.}, ``{Study of the decay
  $\phi \to \eta \pi^0 \gamma$ with the KLOE detector}'',
  \href{http://dx.doi.org/10.1016/S0370-2693(02)01821-X}{{\em Phys. Lett. B}
  {\bfseries 536} (2002) 209--216},
  \href{http://arxiv.org/abs/hep-ex/0204012}{{\ttfamily arXiv:hep-ex/0204012}}.

\bibitem{KLOE:2006vmv}
{\bfseries KLOE} Collaboration, F.~Ambrosino {\em et~al.}, ``{Dalitz plot
  analysis of $e^+ e^- \to \pi^0 \pi^0 \gamma$ events at $\sqrt{s}$
  approximately M($\phi$) with the KLOE detector}'',
  \href{http://dx.doi.org/10.1140/epjc/s10052-006-0157-7}{{\em Eur. Phys. J. C}
  {\bfseries 49} (2007) 473--488},
  \href{http://arxiv.org/abs/hep-ex/0609009}{{\ttfamily arXiv:hep-ex/0609009}}.

\bibitem{BESIII:2015rug}
{\bfseries BESIII} Collaboration, M.~Ablikim {\em et~al.}, ``{Amplitude
  analysis of the \ensuremath{\pi}$^0$\ensuremath{\pi}$^0$ system produced in
  radiative J/\ensuremath{\psi} decays}'',
  \href{http://dx.doi.org/10.1103/PhysRevD.92.052003}{{\em Phys. Rev. D}
  {\bfseries 92} (2015) 052003},
  \href{http://arxiv.org/abs/1506.00546}{{\ttfamily arXiv:1506.00546
  [hep-ex]}}. [Erratum: Phys.Rev.D 93, 039906 (2016)].

\bibitem{E791:2000lzz}
{\bfseries E791} Collaboration, E.~M. Aitala {\em et~al.}, ``{Study of the
  ${\rm D_s^+} \rightarrow \pi^-\pi^+\pi^+$ decay and measurement of f(0)
  masses and widths}'',
  \href{http://dx.doi.org/10.1103/PhysRevLett.86.765}{{\em Phys. Rev. Lett.}
  {\bfseries 86} (2001) 765--769},
  \href{http://arxiv.org/abs/hep-ex/0007027}{{\ttfamily arXiv:hep-ex/0007027}}.

\bibitem{LHCb:2014ooi}
{\bfseries LHCb} Collaboration, R.~Aaij {\em et~al.}, ``{Measurement of
  resonant and CP components in $\bar{B}_s^0\to J/\psi\pi^+\pi^-$ decays}'',
  \href{http://dx.doi.org/10.1103/PhysRevD.89.092006}{{\em Phys. Rev. D}
  {\bfseries 89} (2014) 092006},
  \href{http://arxiv.org/abs/1402.6248}{{\ttfamily arXiv:1402.6248 [hep-ex]}}.

\bibitem{LHCb:2014vbo}
{\bfseries LHCb} Collaboration, R.~Aaij {\em et~al.}, ``{Measurement of the
  resonant and CP components in $\overline{B}^0\to J/\psi \pi^+\pi^-$
  decays}'', \href{http://dx.doi.org/10.1103/PhysRevD.90.012003}{{\em Phys.
  Rev. D} {\bfseries 90} (2014) 012003},
  \href{http://arxiv.org/abs/1404.5673}{{\ttfamily arXiv:1404.5673 [hep-ex]}}.

\bibitem{Daub:2015xja}
J.~T. Daub, C.~Hanhart, and B.~Kubis, ``{A model-independent analysis of
  final-state interactions in $ {\overline{B}}_{d/s}^0\to J/\psi \pi \pi $}'',
  \href{http://dx.doi.org/10.1007/JHEP02(2016)009}{{\em JHEP} {\bfseries 02}
  (2016) 009}, \href{http://arxiv.org/abs/1508.06841}{{\ttfamily
  arXiv:1508.06841 [hep-ph]}}.

\bibitem{Janssen:1994wn}
G.~Janssen, B.~C. Pearce, K.~Holinde, and J.~Speth, ``{On the structure of the
  scalar mesons f0 (975) and a0 (980)}'',
  \href{http://dx.doi.org/10.1103/PhysRevD.52.2690}{{\em Phys. Rev. D}
  {\bfseries 52} (1995) 2690--2700},
  \href{http://arxiv.org/abs/nucl-th/9411021}{{\ttfamily
  arXiv:nucl-th/9411021}}.

\bibitem{Weinstein:1990gu}
J.~D. Weinstein and N.~Isgur, ``{K anti-K Molecules}'',
  \href{http://dx.doi.org/10.1103/PhysRevD.41.2236}{{\em Phys. Rev. D}
  {\bfseries 41} (1990) 2236}.

\bibitem{Xiao:2019lrj}
C.~W. Xiao, U.~G. Mei\ss{}ner, and J.~A. Oller, ``{Investigation of $J/\psi \to
  \gamma\, \pi^0 \eta (\pi^+\pi^-, \pi^0\pi^0)$ radiative decays including
  final-state interactions}'',
  \href{http://dx.doi.org/10.1140/epja/s10050-020-00025-y}{{\em Eur. Phys. J.
  A} {\bfseries 56} (2020) 23},
  \href{http://arxiv.org/abs/1907.09072}{{\ttfamily arXiv:1907.09072
  [hep-ph]}}.

\bibitem{Maiani:2006ia}
L.~Maiani, A.~D. Polosa, V.~Riquer, and C.~A. Salgado, ``{Counting valence
  quarks at RHIC and LHC}'',
  \href{http://dx.doi.org/10.1016/j.physletb.2006.11.072}{{\em Phys. Lett. B}
  {\bfseries 645} (2007) 138--145},
  \href{http://arxiv.org/abs/hep-ph/0606217}{{\ttfamily arXiv:hep-ph/0606217}}.

\bibitem{ExHIC:2017smd}
{\bfseries ExHIC} Collaboration, S.~Cho {\em et~al.}, ``{Exotic hadrons from
  heavy ion collisions}'',
  \href{http://dx.doi.org/10.1016/j.ppnp.2017.02.002}{{\em Prog. Part. Nucl.
  Phys.} {\bfseries 95} (2017) 279--322},
  \href{http://arxiv.org/abs/1702.00486}{{\ttfamily arXiv:1702.00486
  [nucl-th]}}.

\bibitem{Gu:2019oyz}
A.~Gu, T.~Edmonds, J.~Zhao, and F.~Wang, ``{Elliptical flow coalescence to
  identify the $f_{0}$(980) content}'',
  \href{http://dx.doi.org/10.1103/PhysRevC.101.024908}{{\em Phys. Rev. C}
  {\bfseries 101} (2020) 024908},
  \href{http://arxiv.org/abs/1902.07152}{{\ttfamily arXiv:1902.07152
  [nucl-ex]}}.

\bibitem{Bazavov:2018mes}
{\bfseries HotQCD} Collaboration, A.~Bazavov {\em et~al.}, ``{Chiral crossover
  in QCD at zero and non-zero chemical potentials}'',
  \href{http://dx.doi.org/10.1016/j.physletb.2019.05.013}{{\em Phys. Lett. B}
  {\bfseries 795} (2019) 15--21},
  \href{http://arxiv.org/abs/1812.08235}{{\ttfamily arXiv:1812.08235
  [hep-lat]}}.

\bibitem{ALICE:2011dyt}
{\bfseries ALICE} Collaboration, K.~Aamodt {\em et~al.}, ``{Two-pion
  Bose-Einstein correlations in central Pb-Pb collisions at
  $\sqrt{{s}_{\rm{NN}}} =$ 2.76 TeV}'',
  \href{http://dx.doi.org/10.1016/j.physletb.2010.12.053}{{\em Phys. Lett. B}
  {\bfseries 696} (2011) 328--337},
  \href{http://arxiv.org/abs/1012.4035}{{\ttfamily arXiv:1012.4035 [nucl-ex]}}.

\bibitem{ALICE:2018qdv}
{\bfseries ALICE} Collaboration, S.~Acharya {\em et~al.}, ``{Production of the
  $\rho$(770)${^{0}}$ meson in pp and Pb-Pb collisions at $\sqrt{s_{\rm NN}}$ =
  2.76 TeV}'', \href{http://dx.doi.org/10.1103/PhysRevC.99.064901}{{\em Phys.
  Rev. C} {\bfseries 99} (2019) 064901},
  \href{http://arxiv.org/abs/1805.04365}{{\ttfamily arXiv:1805.04365
  [nucl-ex]}}.

\bibitem{ALICE:2019xyr}
{\bfseries ALICE} Collaboration, S.~Acharya {\em et~al.}, ``{Evidence of
  rescattering effect in Pb--Pb collisions at the LHC through production of
  $\rm{K}^{*}(892)^{0}$ and $\phi(1020)$ mesons}'',
  \href{http://dx.doi.org/10.1016/j.physletb.2020.135225}{{\em Phys. Lett. B}
  {\bfseries 802} (2020) 135225},
  \href{http://arxiv.org/abs/1910.14419}{{\ttfamily arXiv:1910.14419
  [nucl-ex]}}.

\bibitem{ALICE:2018ewo}
{\bfseries ALICE} Collaboration, S.~Acharya {\em et~al.}, ``{Suppression of
  $\Lambda(1520)$ resonance production in central Pb-Pb collisions at
  $\sqrt{s_{\rm NN}}$ = 2.76 TeV}'',
  \href{http://dx.doi.org/10.1103/PhysRevC.99.024905}{{\em Phys. Rev. C}
  {\bfseries 99} (2019) 024905},
  \href{http://arxiv.org/abs/1805.04361}{{\ttfamily arXiv:1805.04361
  [nucl-ex]}}.

\bibitem{Fries:2003vb}
R.~J. Fries, B.~Muller, C.~Nonaka, and S.~A. Bass, ``{Hadronization in heavy
  ion collisions: Recombination and fragmentation of partons}'',
  \href{http://dx.doi.org/10.1103/PhysRevLett.90.202303}{{\em Phys. Rev. Lett.}
  {\bfseries 90} (2003) 202303},
  \href{http://arxiv.org/abs/nucl-th/0301087}{{\ttfamily
  arXiv:nucl-th/0301087}}.

\bibitem{Minissale:2015zwa}
V.~Minissale, F.~Scardina, and V.~Greco, ``{Hadrons from coalescence plus
  fragmentation in AA collisions at energies available at the BNL Relativistic
  Heavy Ion Collider to the CERN Large Hadron Collider}'',
  \href{http://dx.doi.org/10.1103/PhysRevC.92.054904}{{\em Phys. Rev. C}
  {\bfseries 92} (2015) 054904},
  \href{http://arxiv.org/abs/1502.06213}{{\ttfamily arXiv:1502.06213
  [nucl-th]}}.

\bibitem{Plumari:2017ntm}
S.~Plumari, V.~Minissale, S.~K. Das, G.~Coci, and V.~Greco, ``{Charmed Hadrons
  from Coalescence plus Fragmentation in relativistic nucleus-nucleus
  collisions at RHIC and LHC}'',
  \href{http://dx.doi.org/10.1140/epjc/s10052-018-5828-7}{{\em Eur. Phys. J. C}
  {\bfseries 78} (2018) 348}, \href{http://arxiv.org/abs/1712.00730}{{\ttfamily
  arXiv:1712.00730 [hep-ph]}}.

\bibitem{ALICE:2014sbx}
{\bfseries ALICE} Collaboration, B.~B. Abelev {\em et~al.}, ``{Performance of
  the ALICE Experiment at the CERN LHC}'',
  \href{http://dx.doi.org/10.1142/S0217751X14300440}{{\em Int. J. Mod. Phys. A}
  {\bfseries 29} (2014) 1430044},
  \href{http://arxiv.org/abs/1402.4476}{{\ttfamily arXiv:1402.4476 [nucl-ex]}}.

\bibitem{Aamodt:2008zz}
{\bfseries ALICE} Collaboration, K.~Aamodt {\em et~al.}, ``{The ALICE
  experiment at the CERN LHC}'',
  \href{http://dx.doi.org/10.1088/1748-0221/3/08/S08002}{{\em JINST} {\bfseries
  3} (2008) S08002}.

\bibitem{ALICE:2013axi}
{\bfseries ALICE} Collaboration, E.~Abbas {\em et~al.}, ``{Performance of the
  ALICE VZERO system}'',
  \href{http://dx.doi.org/10.1088/1748-0221/8/10/P10016}{{\em JINST} {\bfseries
  8} (2013) P10016}, \href{http://arxiv.org/abs/1306.3130}{{\ttfamily
  arXiv:1306.3130 [nucl-ex]}}.

\bibitem{STAR:2002caw}
{\bfseries STAR} Collaboration, C.~Adler {\em et~al.}, ``{Coherent $\rho^0$
  production in ultraperipheral heavy ion collisions}'',
  \href{http://dx.doi.org/10.1103/PhysRevLett.89.272302}{{\em Phys. Rev. Lett.}
  {\bfseries 89} (2002) 272302},
  \href{http://arxiv.org/abs/nucl-ex/0206004}{{\ttfamily
  arXiv:nucl-ex/0206004}}.

\bibitem{ALICE:2015nbw}
{\bfseries ALICE} Collaboration, J.~Adam {\em et~al.}, ``{Coherent
  \ensuremath{\rho}$^{0}$ photoproduction in ultra-peripheral Pb-Pb collisions
  at $ \sqrt{s_{\mathrm{NN}}}=2.76 $ TeV}'',
  \href{http://dx.doi.org/10.1007/JHEP09(2015)095}{{\em JHEP} {\bfseries 09}
  (2015) 095}, \href{http://arxiv.org/abs/1503.09177}{{\ttfamily
  arXiv:1503.09177 [nucl-ex]}}.

\bibitem{Stone:2013eaa}
S.~Stone and L.~Zhang, ``{Use of $B\to J/\psi f_0$ decays to discern the $q
  \bar{q}$ or tetraquark nature of scalar mesons}'',
  \href{http://dx.doi.org/10.1103/PhysRevLett.111.062001}{{\em Phys. Rev.
  Lett.} {\bfseries 111} (2013) 062001}.

\bibitem{ALICE:2016mfm}
{\bfseries ALICE} Collaboration, B.~B. Abelev {\em et~al.}, ``{ALICE luminosity
  determination for pp collisions at $\sqrt{s}=5$ TeV}'',
  \href{http://dx.doi.org/https://cds.cern.ch/record/2202638}{{\em
  ALICE-PUBLIC-2016-005} (2016) }.

\bibitem{Loizides:2017ack}
C.~Loizides, J.~Kamin, and D.~d'Enterria, ``{Improved Monte Carlo Glauber
  predictions at present and future nuclear colliders}'',
  \href{http://dx.doi.org/10.1103/PhysRevC.97.054910,
  10.1103/PhysRevC.99.019901}{{\em Phys. Rev.} {\bfseries C97} (2018) 054910},
  \href{http://arxiv.org/abs/1710.07098}{{\ttfamily arXiv:1710.07098
  [nucl-ex]}}.
[Erratum: Phys. Rev.C99,no.1,019901(2019)].
%%CITATION = ARXIV:1710.07098;%%.

\bibitem{Skands:2014pea}
P.~Skands, S.~Carrazza, and J.~Rojo, ``{Tuning PYTHIA 8.1: the Monash 2013
  Tune}'', \href{http://dx.doi.org/10.1140/epjc/s10052-014-3024-y}{{\em Eur.
  Phys. J. C} {\bfseries 74} (2014) 3024},
  \href{http://arxiv.org/abs/1404.5630}{{\ttfamily arXiv:1404.5630 [hep-ph]}}.

\bibitem{Brun:1994aa}
R.~Brun, F.~Bruyant, F.~Carminati, S.~Giani, M.~Maire, A.~McPherson,
  G.~Patrick, and L.~Urban, ``{GEANT Detector Description and Simulation
  Tool}'', \href{http://dx.doi.org/10.17181/CERN.MUHF.DMJ1}{{\em CERN-W-5013}
  (10, 1994) }.

\bibitem{Acharya:2018qsh}
{\bfseries ALICE} Collaboration, S.~Acharya {\em et~al.}, ``{Transverse
  momentum spectra and nuclear modification factors of charged particles in pp,
  p-Pb and Pb-Pb collisions at the LHC}'',
  \href{http://dx.doi.org/10.1007/JHEP11(2018)013}{{\em JHEP} {\bfseries 11}
  (2018) 013}, \href{http://arxiv.org/abs/1802.09145}{{\ttfamily
  arXiv:1802.09145 [nucl-ex]}}.

\bibitem{Adam:2015bg}
{\bfseries ALICE} Collaboration, J.~Adam {\em et~al.}, ``{Measurement of pion,
  kaon and proton production in proton\textendash{}proton collisions at
  $\sqrt{s} = 7$ TeV}'',
  \href{http://dx.doi.org/10.1140/epjc/s10052-015-3422-9}{{\em Eur. Phys. J. C}
  {\bfseries 75} (2015) 226}, \href{http://arxiv.org/abs/1504.00024}{{\ttfamily
  arXiv:1504.00024 [nucl-ex]}}.

\bibitem{Bellm:2015jjp}
J.~Bellm {\em et~al.}, ``{Herwig 7.0/Herwig++ 3.0 release note}'',
  \href{http://dx.doi.org/10.1140/epjc/s10052-016-4018-8}{{\em Eur. Phys. J. C}
  {\bfseries 76} (2016) 196}, \href{http://arxiv.org/abs/1512.01178}{{\ttfamily
  arXiv:1512.01178 [hep-ph]}}.

\bibitem{Bahr:2008pv}
M.~Bahr {\em et~al.}, ``{Herwig++ Physics and Manual}'',
  \href{http://dx.doi.org/10.1140/epjc/s10052-008-0798-9}{{\em Eur. Phys. J. C}
  {\bfseries 58} (2008) 639--707},
  \href{http://arxiv.org/abs/0803.0883}{{\ttfamily arXiv:0803.0883 [hep-ph]}}.

\bibitem{Lin:2004en}
Z.-W. Lin, C.~M. Ko, B.-A. Li, B.~Zhang, and S.~Pal, ``{A Multi-phase transport
  model for relativistic heavy ion collisions}'',
  \href{http://dx.doi.org/10.1103/PhysRevC.72.064901}{{\em Phys. Rev. C}
  {\bfseries 72} (2005) 064901},
  \href{http://arxiv.org/abs/nucl-th/0411110}{{\ttfamily
  arXiv:nucl-th/0411110}}.

\bibitem{Li:1995pra}
B.-A. Li and C.~M. Ko, ``{Formation of superdense hadronic matter in
  high-energy heavy ion collisions}'',
  \href{http://dx.doi.org/10.1103/PhysRevC.52.2037}{{\em Phys. Rev. C}
  {\bfseries 52} (1995) 2037--2063},
  \href{http://arxiv.org/abs/nucl-th/9505016}{{\ttfamily
  arXiv:nucl-th/9505016}}.

\bibitem{Li:2001xh}
B.~Li, A.~T. Sustich, B.~Zhang, and C.~M. Ko, ``{Studies of superdense hadronic
  matter in a relativistic transport model}'',
  \href{http://dx.doi.org/10.1142/S0218301301000575}{{\em Int. J. Mod. Phys. E}
  {\bfseries 10} (2001) 267--352}.

\bibitem{Wang:1991hta}
X.-N. Wang and M.~Gyulassy, ``{HIJING: A Monte Carlo model for multiple jet
  production in p p, p A and A A collisions}'',
  \href{http://dx.doi.org/10.1103/PhysRevD.44.3501}{{\em Phys. Rev. D}
  {\bfseries 44} (1991) 3501--3516}.

\bibitem{Zhang:1997ej}
B.~Zhang, ``{ZPC 1.0.1: A Parton cascade for ultrarelativistic heavy ion
  collisions}'', \href{http://dx.doi.org/10.1016/S0010-4655(98)00010-1}{{\em
  Comput. Phys. Commun.} {\bfseries 109} (1998) 193--206},
  \href{http://arxiv.org/abs/nucl-th/9709009}{{\ttfamily
  arXiv:nucl-th/9709009}}.

\bibitem{Andersson:1983jt}
B.~Andersson, G.~Gustafson, and B.~Soderberg, ``{A General Model for Jet
  Fragmentation}'', \href{http://dx.doi.org/10.1007/BF01407824}{{\em Z. Phys.
  C} {\bfseries 20} (1983) 317}.

\bibitem{Andersson:1983ia}
B.~Andersson, G.~Gustafson, G.~Ingelman, and T.~Sj{\"o}strand, ``{Parton
  Fragmentation and String Dynamics}'',
  \href{http://dx.doi.org/10.1016/0370-1573(83)90080-7}{{\em Phys. Rept.}
  {\bfseries 97} (1983) 31--145}.

\bibitem{Sjostrand:1993yb}
T.~Sj{\"{o}}strand, ``{High-energy physics event generation with PYTHIA 5.7 and
  JETSET 7.4}'', \href{http://dx.doi.org/10.1016/0010-4655(94)90132-5}{{\em
  Comput. Phys. Commun.} {\bfseries 82} (1994) 74--90}.

\bibitem{Lin:2001zk}
Z.-w. Lin and C.~M. Ko, ``{Partonic effects on the elliptic flow at RHIC}'',
  \href{http://dx.doi.org/10.1103/PhysRevC.65.034904}{{\em Phys. Rev. C}
  {\bfseries 65} (2002) 034904},
  \href{http://arxiv.org/abs/nucl-th/0108039}{{\ttfamily
  arXiv:nucl-th/0108039}}.

\bibitem{Lebiedowicz:2020bwo}
P.~Lebiedowicz, R.~Maciu\l{}a, and A.~Szczurek, ``{Production of $f_{0}(980)$
  meson at the LHC: Color evaporation versus color-singlet gluon-gluon
  fusion}'', \href{http://dx.doi.org/10.1016/j.physletb.2020.135475}{{\em Phys.
  Lett. B} {\bfseries 806} (2020) 135475},
  \href{http://arxiv.org/abs/2003.08200}{{\ttfamily arXiv:2003.08200
  [hep-ph]}}.

\bibitem{ALICE:2019hno}
{\bfseries ALICE} Collaboration, S.~Acharya {\em et~al.}, ``{Production of
  charged pions, kaons, and (anti-)protons in Pb--Pb and inelastic pp
  collisions at $\sqrt {s_{\mathrm{NN}}}$ = 5.02 TeV}'',
  \href{http://dx.doi.org/10.1103/PhysRevC.101.044907}{{\em Phys. Rev. C}
  {\bfseries 101} (2020) 044907},
  \href{http://arxiv.org/abs/1910.07678}{{\ttfamily arXiv:1910.07678
  [nucl-ex]}}.

\bibitem{Hofmann:1988gy}
W.~Hofmann, ``{Particle Composition in Hadronic Jets in $e^+ e^-$
  Annihilation}'',
  \href{http://dx.doi.org/10.1146/annurev.ns.38.120188.001431}{{\em Ann. Rev.
  Nucl. Part. Sci.} {\bfseries 38} (1988) 279--322}.

\bibitem{Chliapnikov:1999qi}
P.~V. Chliapnikov, ``{Hyperfine Splitting in Light-Flavour Hadron Production at
  LEP}'', \href{http://dx.doi.org/10.1016/S0370-2693(99)00910-7}{{\em Phys.
  Lett. B} {\bfseries 462} (1999) 341--353}.

\bibitem{Aguilar-Benitez:1991hzq}
M.~Aguilar-Benitez {\em et~al.}, ``{Inclusive particle production in 400-GeV/c
  pp interactions}'', \href{http://dx.doi.org/10.1007/BF01551452}{{\em Z. Phys.
  C} {\bfseries 50} (1991) 405--426}.

\bibitem{Becattini:2008tx}
F.~Becattini, P.~Castorina, J.~Manninen, and H.~Satz, ``{The Thermal Production
  of Strange and Non-Strange Hadrons in $e^+ e^-$ Collisions}'',
  \href{http://dx.doi.org/10.1140/epjc/s10052-008-0671-x}{{\em Eur. Phys. J. C}
  {\bfseries 56} (2008) 493--510},
  \href{http://arxiv.org/abs/0805.0964}{{\ttfamily arXiv:0805.0964 [hep-ph]}}.

\bibitem{Becattini:1997rv}
F.~Becattini and U.~W. Heinz, ``{Thermal hadron production in p p and p anti-p
  collisions}'', \href{http://dx.doi.org/10.1007/s002880050551}{{\em Z. Phys.
  C} {\bfseries 76} (1997) 269--286},
  \href{http://arxiv.org/abs/hep-ph/9702274}{{\ttfamily arXiv:hep-ph/9702274}}.
  [Erratum: Z.Phys.C 76, 578 (1997)].

\bibitem{Andronic:2017pug}
A.~Andronic, P.~Braun-Munzinger, K.~Redlich, and J.~Stachel, ``{Decoding the
  phase structure of QCD via particle production at high energy}'',
  \href{http://dx.doi.org/10.1038/s41586-018-0491-6}{{\em Nature} {\bfseries
  561} (2018) 321--330}, \href{http://arxiv.org/abs/1710.09425}{{\ttfamily
  arXiv:1710.09425 [nucl-th]}}.

\bibitem{Vovchenko:2019kes}
V.~Vovchenko, B.~D\"onigus, and H.~Stoecker, ``{Canonical statistical model
  analysis of p-p , p -Pb, and Pb-Pb collisions at energies available at the
  CERN Large Hadron Collider}'',
  \href{http://dx.doi.org/10.1103/PhysRevC.100.054906}{{\em Phys. Rev. C}
  {\bfseries 100} (2019) 054906},
  \href{http://arxiv.org/abs/1906.03145}{{\ttfamily arXiv:1906.03145
  [hep-ph]}}.

\bibitem{ALICE:2014jbq}
{\bfseries ALICE} Collaboration, B.~B. Abelev {\em et~al.},
  ``{$\rm{K}^{*}(892)^{0}$ and $\phi(1020)$ production in Pb-Pb collisions at
  $\sqrt{s_{\mathrm{NN}}}$ = 2.76 TeV}'',
  \href{http://dx.doi.org/10.1103/PhysRevC.91.024609}{{\em Phys. Rev. C}
  {\bfseries 91} (2015) 024609},
  \href{http://arxiv.org/abs/1404.0495}{{\ttfamily arXiv:1404.0495 [nucl-ex]}}.

\end{thebibliography}\endgroup
